\documentclass[fleqn,usenatbib]{mnras}

\usepackage{newtxtext,newtxmath}
\usepackage{amsmath}
\usepackage{mathtools}
\usepackage{graphicx}
\usepackage{here}
\usepackage[T1]{fontenc}

\DeclareRobustCommand{\VAN}[3]{#2}
\let\VANthebibliography\thebibliography
\def\thebibliography{\DeclareRobustCommand{\VAN}[3]{##3}\VANthebibliography}

\newcommand{\be}{\begin{equation}}
\newcommand{\ee}{\end{equation}}
\newcommand{\bea}{\begin{eqnarray}}
\newcommand{\eea}{\end{eqnarray}}
\newcommand{\nn}{\nonumber}

\newcommand{\bi}{\begin{itemize}}
\newcommand{\ei}{\end{itemize}}

\usepackage{graphicx}	
\usepackage{amsmath}	

\title[Expanding Fireball in Magnetar Bursts \& FRB]{Expanding Fireball in Magnetar Bursts and Fast Radio Bursts}

\author[Wada \& Ioka]{
Tomoki Wada,$^{1}$\thanks{E-mail: twada@icrr.u-tokyo.ac.jp}
and Kunihito Ioka,$^{2}$
\\
$^{1}$Institute for Cosmic Ray Research, The University of Tokyo, 5-1-5 Kashiwanoha, Kashiwa, Chiba 277-8582, Japan\\
$^{2}$Center for Gravitational Physics and Quantum Information,
Yukawa Institute for Theoretical Physics, Kyoto University, Kyoto, 606-8502, Japan
}

\date{Accepted XXX. Received YYY; in original form ZZZ}

\pubyear{2022}

\begin{document}
\label{firstpage}
\pagerange{\pageref{firstpage}--\pageref{lastpage}}
\maketitle

\begin{abstract}
A fireball of radiation plasma created near the surface of a neutron star (NS) expands under its own pressure along magnetic field lines, and produces photon emission and relativistic matter outflow. 
We comprehensively classify the expanding fireball evolution into five cases and obtain the photospheric luminosity and the kinetic energy of the outflow,
taking into account key processes; lateral diffusion of photons escaping from a magnetic flux tube, effects of strong magnetic field, baryon loading from the NS surface, and radiative acceleration via cyclotron resonant scattering, some of which have not been considered in the context of gamma-ray bursts.
Applying our model to magnetar bursts with fast radio bursts (FRBs), in particular the X-ray short bursts from SGR 1935+2154
associated with the Galactic FRB 20200428A, we show that the burst radiation can accelerate the outflow to high Lorentz factor with
sufficient energy to power FRBs.
\end{abstract}



\section{Introduction}
Magnetars are neutron stars that have strong magnetic field of about $10^{14}$--$10^{15}\,{\rm G}$.
They show diverse flaring activities mainly in X-ray band, short bursts ($\sim 10^{36}$--$10^{43}$ erg s$^{-1}$) and giant flares ($\sim 10^{44}$--$10^{47}$ erg s$^{-1}$), over the super-Eddington luminosity
\citep[][for a review]{KasBel2017,EnoKis2019_neutronstar}.
During outbursts, the flux of the short bursts becomes higher than usual, and the short bursts occur more frequently.
These bursts are considered to be triggered by a sudden release of magnetic energy of the magnetars near the surface where the magnetic energy is highest \citep[][]{ThoDun1995,ThoDun1996,ThoLyu2002}.
The released thermal energy immediately creates electron-positron pairs, forming an optically thick fireball.
If the magnetic energy is higher than the radiative energy of the fireball, the strong magnetic field
confines the fireball in closed magnetic field lines, and a trapped fireball is formed.
The photons diffuse out of the trapped fireball. 
The trapped fireball formed in a giant flare is considered to be the origin of its pulsating tail, which is the slow decay of the luminosity after the initial spike and flicker with the same period as the spin of the magnetar \citep{ThoDun1995}.
In the trapped fireball, the typical energy of the escaping photons is $\sim10\,{\rm keV}$ which shows good agreement with the observation of the short bursts and the pulsating tail of the giant flares.

Fast radio bursts (FRBs) are bright radio transients whose duration are a few milliseconds \citep{LorBai2007,Tho2013,PetBar2016}.
Their high brightness temperature, $\sim 10^{34}\,{\rm K}$, implies that the radio wave is emitted by a coherent radiation process although the concrete emission mechanism and the origin are still unclear \citep[][for review]{Kat2018,PlaWel2019,CorCha2019_FRB,Zha2020_mechanism,Lyu2021_emission,PetHes2021_FRB}.
Regardless of their origin, these bursts can be useful probes for studying cosmology \citep[e.g.,][]{Iok2003,Ino2004,TakIok2021,ShiTak2022}.
In FRBs, some are reported to be repeating sources \citep{Spi2014,Spi2016,CHIME2019_2nd,CHIME2019_8,Bha2021}, which suggest that at least some FRBs have noncatastrophic origins.
Almost all FRBs have extragalactic origin except one Galactic FRB, FRB 20200428A \citep{Boc2020,CHIME2020_200428}.
FRB 20200428A is an exceptional burst from a Galactic magnetar, SGR1935+2154, and is associated with hard short bursts in X-ray band \citep{Mer2020,Li2021,Rid2021,Tav2021,LiGe2022}.
When FRB 20200428A is detected, SGR1935+2154 is in its outburst and two of the short bursts are detected with the FRB \citep{YouBar2021,CaiXue2022}.
This event suggests that some FRBs are from magnetars and 
most likely
associated with X-ray flares \citep{Kat2016,Kat2020}.
The isotropic luminosity of FRB 20200428A is $\sim 10^{38}\,{\rm erg\,s^{-1}}$, and that of the short burst is $\sim 10^{41}\,{\rm erg\,s^{-1}}$.
The cutoff energy of the X-ray bursts associated with FRB 20200428A is about $80\,{\rm keV}$, which is higher than the typical energy realized in the trapped fireball, $\sim 10\,{\rm keV}$.

This observed high cutoff energy of the X-ray burst associated with FRB 20200428A, $\sim80\,{\rm keV}$, can be realized in an expanding fireball \citep{Iok2020}, which is the focus of this paper.\footnote{
The inverse Compton scattering by the outflow of the X-ray bursts
could be also important
\citep{YanZha2021,ZhaZha2021,YamKas2022}.}
In the context of gamma-ray bursts, a spherically symmetric expanding fireball has been investigated \citep[e.g.,][for review]{Pir1999,Zha2018G}.
For a spherically symmetric fireball of purely pair plasma, as it expands, the fluid is accelerated to a relativistic speed, the temperature in the comoving frame of the fluid decreases, the number density of pairs in the frame also decreases, and eventually, photons escape from the fireball at a photospheric radius \citep{Goo1986,Pac1986}.
If some amount of baryon is loaded into the fireball, the electrons associated with the baryons make the photospheric radius larger, the acceleration continues longer, and the kinetic luminosity of outflow becomes higher \citep{ShePir1990,MesLag1993,MesRee2000}.
The observed temperature equals the initial temperature of the fireball if the radiative energy density is higher than the rest-mass energy density at the photospheric radius.
Therefore, if the initial temperature of the expanding fireball is $\sim 80\,{\rm keV}$, the observed temperature can also be $\sim 80\,{\rm keV}$.
These expanding fireball is also applied to the giant flares of the magnetars \citep{ThoDun2001,NakPir2005}.

In this paper, we study the evolution of a fireball expanding along open magnetic field lines.\footnote{
The open magnetic field lines here do not necessarily have to be open to infinity. 
If the magnetic field lines are closed on a larger scale than we consider, it will not affect the following discussions.}
In Section~\ref{sec:dynamics}, we consider the dynamics of the fireball along open dipole magnetic field lines.
First, the dynamics of a relativistic fluid flowing along open magnetic field lines is investigated.
Next, taking into account the effect of a strong magnetic field, we estimate the optical depth and diffusion time, and identify where the photons escape from the fireball.
The baryon loading to the fireball is taken into account.
Different from a spherically expanding fireball, photons can diffuse out of the fireball laterally across the magnetic field lines before the fireball becomes optically thin \citep{Iok2020}.
These baryon loading and diffusion are fully investigated for the first time.
The terminal Lorentz factor of the outflow can be enhanced by radiative acceleration via cyclotron resonant scattering, and this effect is also investigated for the first time.
In Section~\ref{sec:cases}, we identify five cases of the escaping process of the photon from the fireball and investigate the outflows from the fireball for all cases depending on the amount of baryon.
In Section~\ref{sec:200428}, we also apply our model to the observed X-ray burst associated with Galactic FRB, FRB 20200428A.
Section~\ref{sec:summary} is devoted to the conclusion and discussion.

In this paper, we normalize the temperature and the magnetic field as 
\begin{equation}
  \frac{k_{\rm B}T}{m_{\rm e}c^2}\to T,\quad \frac{B}{B_{\rm Q}}\to B,
  \label{eq:normalize}
\end{equation}
where $k_{\rm B}$, $m_{\rm e}$, $c$, $B_{\rm Q}=m_{\rm e}^2c^3/(e\hbar)$, and $\hbar$ are the Boltzmann constant, the electron mass, the speed of light, the critical field strength, and the Planck constant divided by $2\pi$.
We use $Q_x=Q/10^{x}$ in cgs units.

\section{Fireball dynamics along Open magnetic field lines}\label{sec:dynamics}
\begin{figure}
\includegraphics[width=\linewidth]{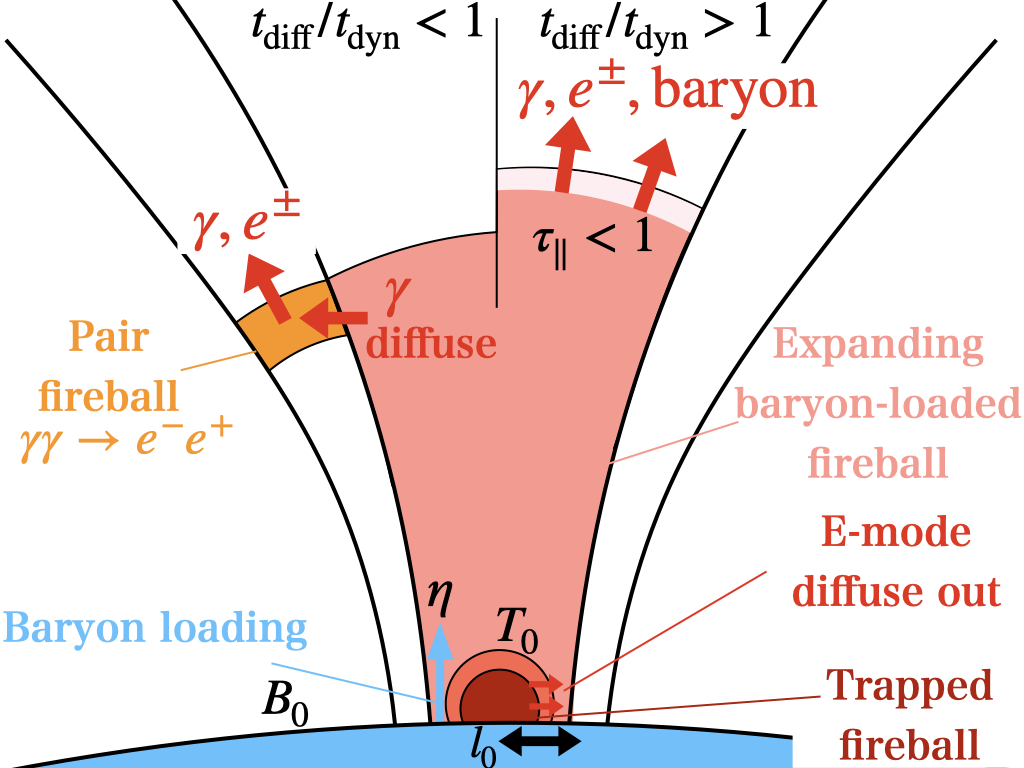}
\caption{
  Schematic diagram of the fireball expanding along open magnetic field lines.
  The expanding fireball releases radiation for X-ray bursts and generates the outflow energy for FRBs.
  At the base of an open magnetic field, 
  a small-scale trapped fireball could be formed
  and diffusively supply electron-positron pairs and radiation to flux tube of open magnetic field lines.
  From the surface of the magnetar, baryon is loaded into the fireball and baryon-loaded fireball expanding along the open field line is created.
  As it flows, the fireball is accelerated to a relativistic speed, the number density decreases, and photons escape from the fireball in two ways.
  If the optical depth in the direction parallel to the magnetic field lines becomes smaller than unity ($\tau_\parallel<1$), the photons escape from the fireball longitudinally
  (the right half of this figure).
  On the contrary, if the diffusion timescale of photons perpendicular to the magnetic field lines becomes shorter than the dynamical timescale ($t_{\rm diff}<t_{\rm dyn}$), the photons diffuse out laterally from the fireball via diffusion (the left half of this figure).
  In this case, the fireball may expand laterally due to the pair creation if the temperature of the fireball is high enough (the yellow region in the left of this figure).}
\label{fig:overview}
\end{figure}

We consider the dynamics of a fireball along open dipole magnetic field in a magnetar (Figure~\ref{fig:overview}).
The open magnetic field has a larger scale than the expanding fireball.
This fireball is formed if crustal shear oscillation or magnetic reconnection release thermal energy in a small region and electron-positron pairs with a high optical depth are created.
If this process occurs in closed magnetic field lines, a trapped fireball would be formed \citep{ThoDun1995}.
However, the trapped fireball cannot explain the high-temperature X-ray burst associated with Galactic FRB, FRB 20200428A \citep{,Mer2020,Li2021,Rid2021,Tav2021}.
This is because, at a high temperature, the optical depth is so high that the photons cannot escape from the trapped fireball \citep{Iok2020}.
Therefore, in order to realize the high temperature, photons have to be emitted from a relativistic outflow like a spherically symmetric fireball case.
That is why we consider a fireball along open magnetic field lines.

At the bottom of the open magnetic field, a small-scale closed magnetic field would exist and a small-scale trapped fireball can be formed.
Because the magnetic field is strong, the scattering cross section for an E-mode photon\footnote{
In magnetized plasma, two eigen modes for waves are defined, E-mode (extraordinary) and O-mode (ordinary).
The electric field of an E-mode photon is perpendicular to the plane that contains the wavenumber vector and the background magnetic field vector.
The electric field of an O-mode photon is on that plane.}
is suppressed \citep{CanLod1971,Her1979,Meszaros1992}.
Thus, E-mode photons come out more easily than O-mode photons from the trapped fireball \citep{ThoDun1995}.
Because the trapped fireball is optically thick, the E-mode photons diffuse out and produce electron-positron pairs outside of the trapped fireball and in the flux tube of the open magnetic field lines.
As a result, the fireball expands laterally with respect to the magnetic field lines, and eventually, the pair plasma would flow along the large-scale open magnetic field lines \citep[e.g.,][]{ThoDun2001,Iok2020,YanZha2021,ZhaZha2021}.

Because the flux of the escaping photon is higher than the Eddington flux, the surface of the star is heated, and baryons are loaded in the fireball \citep{ThoDun1995}.
In this way, the baryon-loaded fireball may be created and it spreads along the open magnetic field lines.

In the open magnetic field lines, the comoving temperature of the fireball is $T$, the lateral size of the fireball is $\ell$, the comoving rest-mass density of the baryon is $\rho$, and the comoving energy density of the radiation is
\begin{equation}
  e=\left(\frac{\pi^2m_{\rm e}^4c^5}{15\hbar^3}\right)T^4,\label{eq:at4}
\end{equation}
(see Figure~\ref{fig:overview} and Equation (\ref{eq:normalize})).
$\Gamma$ is the Lorentz factor of the fireball in the lab frame and we assume that the initial Lorentz factor is 1.
The magnetic field strength on the magnetar surface is $B_0$, and the radius of the neutron star is $r_0$.
The subscript 0 of $T$, $\ell$, $\rho$, $e$, and $\Gamma$, represents that it is the initial value at $r_0$.
The dimensionless entropy is
\begin{equation}
   \eta=\frac{e_0}{\rho_0c^2}.\label{eq:eta}
\end{equation}
These quantities are normalized as 
\begin{equation}
\tilde{r}=r/r_0,\quad \tilde{\rho}=\rho/\rho_0,\quad \tilde{e}=e/e_0,\quad \tilde{T}=T/T_0,
\end{equation}
and with this normalization, the lateral size of the fireball is
\begin{equation}
  \ell=\ell_0\tilde{r}^{3/2}.
\end{equation}
We use polar coordinates, $(r,\vartheta)$, whose origin is at the center of the magnetar.
We take $\vartheta=0$ to coincide with the direction of the magnetic dipole moment.

\subsection{Fluid Approximation}\label{fluid}
\begin{table*}
  \caption{Summary of the evolution of physical quantities and the characteristic radii.
 $\tilde{T}$, $\Gamma$, $\tilde{\rho}$ are the temperature, Lorentz factor, and baryon rest-mass density of the fireball at $\tilde{r}$ (Equations~(\ref{eq:sol_e})--(\ref{eq:sol_rho})).
 We used $\tilde{e}=\tilde{T}^4$ to calculate $\tilde{T}$.
 $\tilde{r}_{\rm S}$ is the saturation radius where the fireball becomes matter-dominated (Equation~(\ref{eq:r_s})).
 $\tilde{r}_{\rm A}$ is the Alfv\'{e}n radius where the magnetic energy density of the dipole field becomes lower than the energy density of the fluid (Equation~(\ref{eq:alfven})). 
 $\tilde{r}_{\rm B}$ is the Landau-level crossing radius where the number density of the pair plasma changes from Equation~(\ref{eq:number_BT}) to Equation~(\ref{eq:number_T}).
 For the radiation-dominated phase and $B\gg B_{\rm Q}$ case, $\tilde{r}_{\rm B}$ does not exist because both $\hbar\omega_{\rm e}(1)$ and $m_{\rm e}c^2T$ have the same dependence on $\tilde{r}$.
 $\tilde{r}_{\rm E}$ is the scattering-suppression radius below which the cross section for an E-mode photons is suppressed.
 }
 \label{tab:radbar_summary}
 \centering
  \begin{tabular}{ccc}
   \hline
     & Radiation-dominated & Matter-dominated \\
   \hline \hline
   $\tilde{T}$ & $\tilde{r}^{-3/2}$ &$\eta^{-1/3}\tilde{r}^{-1}$\\
   $\Gamma$ & $\tilde{r}^{3/2}$ & $\eta$\\
   $\tilde{\rho}$ & $\tilde{r}^{-9/2}$ & $\eta^{-1}\tilde{r}^{-3}$\\
   \hline
   $\tilde{r}_{\rm S}$ & $\eta^{2/3}$ & $\eta^{2/3}$ \\
   \hline
    $\tilde{r}_{\rm A}$ &no solution & $\left(15/8\pi^3\alpha\right)^{1/3}B_0^{2/3}T_0^{-4/3}\eta^{2/3}$\\
    \hline
   $\tilde{r}_{\rm B}$ ($B\gg B_{\rm Q}$) & no solution  & $ 2B_0T_0^{-2}\eta^{2/3}$ \\
   $\tilde{r}_{\rm B}$ ($B\ll B_{\rm Q}$) & $B_{0}^{2 /3}T_{0}^{-2 / 3}$ & $B_{0}^{1/ 2}T_{0}^{-1 / 2} \eta^{1/6}$\\
   \hline
   $\tilde{r}_{\rm E}$ & $\left(4\pi^2/5\right)^{-1/3}T_0^{-2/3}B_0^{2/3}$ & $\left(4\pi^2/5\right)^{-1/4}T_0^{-1/2}B_0^{1/2}\eta^{1/6}$\\
   \hline
  \end{tabular}
\end{table*}

First, we consider the dynamics of fluid moving along open magnetic field lines.
If the fireball is optically thick, photons, electrons, positrons, and baryons are described as a one-component fluid.
The initial conditions to be satisfied are discussed in Section~\ref{sec:initial}, and the decoupling of photons is described in Section~\ref{sec:beaking}.
We use the radial coordinate, $r$, to measure the position of the fluid.
Strictly speaking, each fluid element flows along magnetic field lines and the position of the fluid deviates from the spherically symmetric case.
However, we ignore this deviation.
This approximation breaks
around $\tilde{r}=\theta_0^{-2}$ \citep{ThoDun2001}, where
\begin{equation}
\theta_0=\frac{\ell_0}{r_0},
\end{equation}
and $\theta_0\ll 1$ is assumed.
In this paper, we mainly consider the case of $\theta_0\lesssim 10^{-2}$.
For $\theta_0\lesssim 10^{-2}$ and $T_0\lesssim1 $, the diffusion of pairs (see Section~\ref{sec:beaking}) occurs.
Even for $10^{-2}\lesssim\theta_0\lesssim10^{-1}$, if $T_0\lesssim0.3$ is satisfied, the diffusion of pairs occur and our formulation is valid.

In this paper, we assume the axisymmetric dipole structure of the magnetic field.
For a dipole magnetic field, the strength of the magnetic field is \citep[e.g.,][]{Jackson1975}
\begin{equation}
    B(r)=B_0\tilde{r}^{-3}.
    \label{eq:Br}
\end{equation}
We neglect $\vartheta$ dependence of the magnetic field for simplicity.
The general multipolar cases are shown in Appendix~\ref{sec:multipole}.

We assume that the constant energy injection continues on a timescale longer than the dynamical timescale.
We also assume that entropy generation, such as shock and dissipation of the magnetic field, does not occur during the expansion.
In this case, the flow is steady and adiabatic.

Under these assumptions, the baryon number density flux, the energy density flux, and the entropy flux of the fluid is conserved along the initial magnetic field lines (flux tube).
Since the temperature is usually less than the baryon rest mass energy, we consider the fluid consisting of cold baryon gas and radiation.
These conservation laws are expressed as \citep[e.g.,][]{Pac1986,ThoDun2001}
\begin{eqnarray}
r^{2} \Delta \Omega \,\rho \Gamma \beta &=\text { const. }, \label{eq:baryon_dyn}\\
r^{2} \Delta \Omega\,\left(\rho c^2+\frac{4}{3} e\right) \Gamma^{2} \beta^{2} &=\text { const. }, \label{eq:energy_dyn}\\
r^{2} \Delta \Omega\, e^{3 / 4} \Gamma \beta &=\text { const. }\label{eq:entropy_dyn},
\end{eqnarray}
where $\Delta\Omega$ is the solid angle of the flux tube, $\beta=\sqrt{1-\Gamma^{-2}}$ is the velocity of the fluid  in the lab frame normalized by the speed of light, and we assume that the baryon number density is proportional to its rest-mass density.
For a dipole magnetic field, $r^2\Delta \Omega$ is proportional to $r^{3}$ in the flux tube.
Hereafter, we set $\beta=1$ because the flow is accelerated to a relativistic speed by the strong radiation pressure.

In the radiation-dominated phase where $e\gg\rho c^2$ holds, we approximate $\rho c^2 + 4e/3\sim 4e/3$ in Equation~(\ref{eq:energy_dyn}).
Thus, $e, \Gamma$, and $\rho$ depend on $r$ as,
\begin{eqnarray}
e &\propto& r^{-6},\label{eq:er_rad} \\
\Gamma &\propto& r^{3/2},\label{eq:gamma_rad} \\
\rho &\propto& r^{-9/2}.\label{eq:rho_rad} 
\end{eqnarray}
Each quantity depends on $r$ more strongly than in the spherically symmetric case.
This is due to the rapid lateral expansion of the fluid along the flux tube.
In the radiation-dominated phase, the observed temperature of the fireball,
\begin{equation}
    T_{\rm obs}=\Gamma T,\label{eq:T_obs}
\end{equation}
does not depend on $r$ as in the spherically symmetric case (see Equation~(\ref{eq:at4})).

In the matter-dominated phase where $e\ll\rho c^2$ holds, we approximate $\rho c^2+ 4e/3\sim \rho c^2$ in Equation~(\ref{eq:energy_dyn}).
$e, \Gamma$, and $\rho$ depend on $r$ as,
\begin{eqnarray}
e &\propto& r^{-4}, \label{eq:er_mat}\\
\Gamma &\propto& r^{0}={\rm const.}, \\
\rho &\propto& r^{-3}.\label{eq:rho_mat}
\end{eqnarray}
The transition from the radiation-dominated phase to the matter-dominated phase occurs, where $1=e/\rho c^2=\eta \tilde{r}^{-3/2}$.
We define this saturation radius as,
\begin{equation}
    \tilde{r}_{\rm S}\coloneqq\eta^{2/3}.\label{eq:r_s}
\end{equation}

Matching the solutions in the radiation-dominated phase (Equations~(\ref{eq:er_rad})--(\ref{eq:rho_rad})) and the solutions in the matter-dominated phase (Equations~(\ref{eq:er_mat})--(\ref{eq:rho_mat})) at $\tilde{r}_{\rm S}$, the solutions of Equations~(\ref{eq:baryon_dyn})--(\ref{eq:entropy_dyn}) are
\begin{eqnarray}
\tilde{e}(r)&=&
\begin{dcases}
\tilde{r}^{-6} &({\rm RD})\\
\eta^{-4/3}\tilde{r}^{-4}&({\rm MD}),
\end{dcases}\label{eq:sol_e}\\
\Gamma(r)&=&
\begin{dcases}
\tilde{r}^{3/2}&({\rm RD})\\
\eta&({\rm MD}),
\end{dcases}\label{eq:sol_gamma}\\
\tilde{\rho}(r)&=&
\begin{dcases}
\tilde{r}^{-9/2}&({\rm RD})\\
\eta^{-1}\tilde{r}^{-3}&({\rm MD}),
\end{dcases}\label{eq:sol_rho}
\end{eqnarray}
where RD means radiation-dominated phase ($\tilde{r}<\tilde{r}_{\rm S}$) and MD means matter-dominated phase ($\tilde{r}>\tilde{r}_{\rm S}$).
These dependencies are summarized in Table~\ref{tab:radbar_summary}.
Hereafter, $e(r)$, $\Gamma(r)$, and $\rho(r)$ are defined as the functions in Equations~(\ref{eq:sol_e})--(\ref{eq:sol_rho}).

In the radiation-dominated phase, if the magnetic energy density of the dipole magnetic field is higher than the radiation energy density at the base of the flux tube, the magnetic energy density remains higher than the radiation energy density because both are proportional to $r^{-6}$ (see also Equation~(\ref{eq:maghighrad})).
After the fireball becomes matter-dominated, the rest mass energy density of baryons becomes higher than the magnetic energy density at Alfv\'{e}n radius, $\tilde{r}_{\rm A}$.
From Equations~(\ref{eq:Br}) and (\ref{eq:sol_rho}), $r_{\rm A}$ 
is 
\begin{equation}
    \tilde{r}_{\rm A}=\left(\frac{15}{8\pi^3\alpha}\right)^{1/3}B_0^{2/3}T_0^{-4/3}\eta^{2/3}
    \label{eq:alfven}
\end{equation}
where $\alpha $ is the fine-structure constant.
Above the Alfv\'{e}n radius, the dipole magnetic field no longer confines the fireball in the flux tube.

\subsection{Effects of the Magnetic Field on Pair Number Density and Cross Section}\label{sec:magnetic}
In strong magnetic field, the equilibrium number density of $e^\pm$ and the cross section of Compton scattering for an E-mode photon change.
In this subsection, we give a brief overview of these effects.
We do not take the baryonic component into account in this subsection.
The electrons associated with the baryons are considered in Section~\ref{sec:beaking}.
We consider the case that $m_{\rm e}c^2T<m_{\rm e}c^2$.
We note that the dimensionless entropy, $\eta=e_0/(\rho_0c^2)$, is the ratio of the radiation energy density to the rest mass energy density of the baryonic component, and thus $T<m_{\rm e} c^2$ and $\eta>1$ are compatible.
The chemical potential can be ignored here because of the thermal equilibrium ($\mu_++\mu_-=2\mu_\gamma=0$)
and the charge neutrality ($\mu_+=\mu_-$),
where $\mu_\gamma,\,\mu_+,\,\mu_-$ are the chemical potentials of the photons, the positrons, and the electrons.

\subsubsection{Equilibrium Number Density of Pairs}\label{sec:number}
We consider the equilibrium number density of $e^\pm$ in a strong magnetic field.
The number density varies depending on whether or not the excitation energy of the first Landau level exceeds the temperature.
In a strong magnetic field, the first Landau level is given as \citep{Lan1930,JohLip1949}
\begin{eqnarray}
 \hbar\omega_{\rm e}(1)&=&m_{\rm e}c^2\left(\sqrt{1+2B}-1\right)\nonumber\\
 &=&\begin{dcases}
    m_{\rm e}c^2B &(B\ll 1)\\
    m_{\rm e}c^2\sqrt{2B}&(B\gg 1)
	\end{dcases}.\label{eq:Landau}
\end{eqnarray}

First we consider the case that the temperature is much lower than the electron rest mass energy and much higher than the first Landau level ($m_{\rm e}c^2\gg m_{\rm e}c^2T\gg\hbar \omega_{\rm e}(1)$).
In this case, the higher Landau levels are occupied and the number density is nearly the same as that without the magnetic field.
Thus, the pair number density in thermal equilibrium, $n_\pm$, which is the sum of the number density of the electrons and positrons in thermal equilibrium, is
 \begin{eqnarray}
  n_\pm(T)&=&4\left(\frac{m_{\rm e}^2 c^2}{2 \pi \hbar^{2}}\right)^{3 / 2} T^{3/2} \exp\left(-T^{-1}\right).\label{eq:number_T}
\end{eqnarray}

Next, we consider the case that the temperature is much lower than both the first Landau level and the electron rest mass energy ($m_{\rm e}c^2,\, \hbar\omega_{\rm e}(1)\gg  m_{\rm e}c^2T$).
In this case, almost all electrons are in the lowest Landau level, and the number density is \citep{ThoDun1995}
\begin{eqnarray}
 n_\pm(T,\,B)&=&4\left(\frac{m_{\rm e}^2 c^2}{2 \pi \hbar^{2}}\right)^{3 / 2}BT^{1/2} \exp{\left(-T^{-1}\right)}\nn\\
  &=&n_\pm(T)T^{-1}B.\label{eq:number_BT}
\end{eqnarray}

Even if $m_{\rm e}c^2,\, \hbar\omega_{\rm e}(1)\gg  m_{\rm e}c^2T$ is satisfied initially, as the radius increases, this relation is broken and the inequality, $m_{\rm e}c^2\gg m_{\rm e}c^2T\gg \hbar\omega_{\rm e}(1)$, becomes satisfied in the radiation-dominated phase.
The reason is as follows.
If the magnetic field is stronger than $B_{\rm Q}$, both $\hbar\omega_{\rm e}(1)$ and $m_{\rm e}c^2 T$ are proportional to $r^{-3/2}$ (see Equations~(\ref{eq:Br}), (\ref{eq:er_rad}), and (\ref{eq:Landau})), and the inequality, $m_{\rm e}c^2,\, \hbar\omega_{\rm e}(1)\gg  m_{\rm e}c^2T$, remains satisfied.
However, after the magnetic field becomes weaker than $B_{\rm Q}$, $\hbar\omega_{\rm e}(1)$ becomes proportional to $r^{-3}$ (see Equation~(\ref{eq:Landau})) while $m_{\rm e}c^2T$ is proportional to $r^{-3/2}$.
Thus, at the Landau-level crossing radius, $\tilde{r}_{\rm B}$, $\hbar\omega_{\rm e}(1)$ becomes smaller than $m_{\rm e}c^2T$, and the higher Landau levels begin to be occupied.
Thus, the equilibrium number density changes from Equation~(\ref{eq:number_BT}) to Equation~(\ref{eq:number_T}).
In the matter-dominated phase, a similar thing occurs, and the equilibrium number density changes.
The specific expressions of the Landau-level crossing radius, $\tilde{r}_{\rm B}$, are summarized in Table~\ref{tab:radbar_summary}.

Hereafter, to simplify the notation, we write the density of the positron as $n_+(T,B)$, which equals $n_\pm(T)/2$ in Equation~(\ref{eq:number_T}) if the higher Landau levels are occupied ($\tilde{r}\geq\tilde{r}_{\rm B}$) and equals $n_\pm(T,B)/2$ in Equation~(\ref{eq:number_BT}) if only the lowest Landau level is occupied ($\tilde{r}<\tilde{r}_{\rm B}$).

\subsubsection{Cross Section for E-mode Photons}
We consider the cross section for photons, whose energy is lower than $m_{\rm e}c^2$, scattered by an electron in strong magnetic field.
In this case, the cross section for an E-mode photon is suppressed compared to that of  Thomson scattering while that of an O-mode photon is not.
This is because the motion of an electron in the direction perpendicular to the magnetic field is suppressed.
The Rosseland mean of the cross section is \citep{Meszaros1992,ThoDun1995}
\begin{equation}
 \sigma_{\rm E}(T,\,B)=\frac{4\pi^2}{5}T^2B^{-2}\sigma_{\rm T}\label{eq:crosssection},
\end{equation}
where $\sigma_{\rm T}$ is the Thomson scattering cross section.

The suppression of the cross section for an E-mode photon is the strongest at the base of the magnetic field line, and becomes weaker as the radius increases.
Eventually, $\sigma_{\rm E}$ becomes equal to $\sigma_{\rm T}$, and the photons of both modes are scattered with the Thomson cross section, $\sigma_{\rm T}$.
We define this scattering-suppression radius as $\tilde{r}_{\rm E}$ and the specific expressions are summarized in Table~\ref{tab:radbar_summary}.

Hereafter, we write the cross section as $\sigma(T,B)$, which equals $\sigma_{\rm E}(T,\,B)$ if the suppression for the E-mode photons occurs ($\tilde{r}<\tilde{r}_{\rm E}$) and equals $\sigma_{\rm T}$ if the suppression does not occur ($\tilde{r}\geq \tilde{r}_{\rm E}$).\footnote{
Even if the fireball is optically thin for E-mode photons, it may be optically thick for O-mode photons.
However, once E-mode photons escape from the fireball, O-mode photons also begin to escape (see Section~\ref{sec:pl} for detail).}

\subsubsection{Photon Splitting}
If the timescale of photon splitting in strong magnetic field, $\gamma B\to\gamma\gamma$, is smaller than the dynamical timescale, the photons escaping from the fireball can split into two photons.
However, for $B_0B_{\rm Q}\lesssim10^{15}\,{\rm G}$, the timescale of the photon splitting is longer than the dynamical timescale at the radius where photons escape from the fireball ($\tilde{r}\gtrsim 3$, see Section~\ref{sec:beaking} and Figure~\ref{fig:typical_r}).
Therefore, we neglect this effect in this paper.

\subsection{Initial Condition}\label{sec:initial}
In this paper, we assume the following two initial conditions.
First, the fireball is optically thick at the base of the flux tube.
If the magnetic field is strong, we should consider the suppression of the cross section for the E-mode photons and the Landau level for the electron number density (see Section~\ref{sec:magnetic}).
Depending on the strength of the magnetic field, there are three kinds of initial conditions.
\begin{itemize}
\item[(i)] In the case that the suppression for the E-mode occurs and only the lowest Landau level is occupied, the initial fireball is optically thick if
\begin{eqnarray}
 \tau_{\rm BE0}&\coloneqq& n_\pm(T_0,B_0)\sigma_{\rm E}(T_0,B_0) \ell_0\nn\\
&=&2.4\times10^{11}\,\exp\left(-T_0^{-1}\right)T_0^{5/2}B_0^{-1}\ell_{0,4},
\end{eqnarray}
is higher than unity.
\item[(ii)] In the case that the suppression for the E-mode does not occur and only the lowest Landau level is occupied, the optically-thick condition is
\begin{eqnarray}
  1<\tau_{\rm B0}&\coloneqq& n_\pm(T_0,B_0)\sigma_{\rm T} \ell_0\nn\\
 &=&3.0\times10^{10}\,\exp\left(-T_0^{-1}\right)T_0^{1/2}B_0\ell_{0,4}.
\end{eqnarray}
\item[(iii)] In the case that the suppression for the E-mode photons does not occur and higher Landau levels are occupied, the optically-thick condition is 
\begin{eqnarray} 
  1<\tau_{0}&\coloneqq& n_\pm(T_0)\sigma_{\rm T} \ell_0 \nn\\
&=&3.0\times10^{10}\,\exp\left(-T_0^{-1}\right)T_0^{3/2}\ell_{0,4}.
\end{eqnarray}
\end{itemize}
In all cases, the fireball is optically thick for the parameter range which we are interested in.

Next, we consider the condition that the magnetic energy density is higher than the radiation energy density at the base of the flux tube.
Otherwise, the magnetic field cannot confine the pair plasma.
This condition is expressed as
\begin{eqnarray}
\frac{B_0^2B_{\rm Q}^2}{8\pi}&>&e_0, \nn \\
B_0&>&\left(\frac{8 \pi^{3}}{15} \alpha\right)^{1 / 2}T_0^2\label{eq:maghighrad}\\
&=&0.35T_0^2.\nn
\end{eqnarray}
If this condition is satisfied at the base, it remains satisfied in the radiation-dominated phase.
This is because the radial dependence of both energies are the same, $\propto r^{-6}$, (see, Equations~(\ref{eq:Br}) and (\ref{eq:er_rad})).

\subsection{Decoupling of Photons}\label{sec:beaking}

\begin{table*}
 \caption{Variables in Equations~(\ref{eq:tau_b}) and (\ref{eq:diff_b}).
 The leftmost box shows each case.
 RD means radiation-dominated phase, and MD means matter-dominated phase.
 O-mode means the case where the suppression for the E-mode photons does not occur, and E-mode means the case where it occurs.}
 \label{tab:index_baryon}
 \centering
  \begin{tabular}{c|ccccccc}
   \hline
    & $\tau_{\rm b0}$&$\gamma$ & $\delta$ & $\epsilon$ & $\zeta$\\
   \hline \hline
   RD/O-mode ($\tilde{r}_{\rm S}>\tilde{r}>\tilde{r}_{\rm E}$) & $e_0\sigma_{\rm T} r_0/m_{\rm p}c^2$& $-1$ & $-5$ & $-1$ & $-1$\\
   \hline
   MD/O-mode ($\tilde{r}>\tilde{r}_{\rm S}$, $\tilde{r}_{\rm E}$)& $e_0\sigma_{\rm T} r_0/m_{\rm p}c^2$& $-3$ & $-2$ & $-1$ & $-1$\\
   \hline
   RD/E-mode ($\tilde{r}_{\rm E},\tilde{r}_{\rm S}>\tilde{r}$) & $(4\pi^2T_0^2B_0^{-2}/5)e_0\sigma_{\rm T} r_0/m_{\rm p}c^2$& $-1$ & $-2$ & $-1$ & $2$\\
   \hline
   MD/E-mode ($\tilde{r}_{\rm E}>\tilde{r}>\tilde{r}_{\rm S}$) &$(4\pi^2T_0^2B_0^{-2}/5)e_0\sigma_{\rm T} r_0/m_{\rm p}c^2$ & $-11/3$ & $2$ & $-5/3$ & $3$\\
   \hline
  \end{tabular}
\end{table*}

\begin{table*}
 \caption{Variables in Equations~(\ref{eq:tau_p}) and (\ref{eq:diff_p}).
 The meanings of RD, MD, O-mode, and E-mode are the same as Table~\ref{tab:index_baryon} ("-mode" is omitted for simplicity).
 hL means that the higher Landau states are occupied, and lL means that only the lowest Landau level is occupied.
 In RD, $\eta$ dependence does not appear ($\gamma^\prime=0$ and $\epsilon^\prime=0$) because the number density of the pair plasma is not affected by the baryons.
 $n_\pm(T_0)$ and $n_\pm(T_0,B_0)$ are in Equations~(\ref{eq:number_T}) and (\ref{eq:number_BT}.)
 }
 \label{tab:index_pair}
 \centering
  \begin{tabular}{c|cccccccc}
   \hline
    & $\tau_{\rm \pm0}$&$\gamma^\prime$ & $\delta^\prime$ & $\epsilon^\prime$ & $\zeta^\prime$ & $A$\\
   \hline \hline
   RD/O-mode/lL & $n_\pm(T_0,B_0)\sigma_{\rm T}r_0$& $0$ & $-17/4$ & $0$ & $-1/4$ & $\tilde{r}^{3/2}-1$\\
   \hline
    RD/O-mode/hL & $n_\pm(T_0)\sigma_{\rm T}r_0$& $0$ & $-11/4$ & $0$ & $5/4$ & $\tilde{r}^{3/2}-1$\\
   \hline
   MD/O-mode/lL & $n_\pm(T_0,B_0)\sigma_{\rm T}r_0$ & $-7/6$ & $-5/2$ & $5/6$ & $-3/2$&$\eta^{1/3}\tilde{r}-1$\\
   \hline
    MD/O-mode/hL & $n_\pm(T_0)\sigma_{\rm T}r_0$& $-3/2$ & $-1/2$ & $1/2$ & $1/2$ &$\eta^{1/3}\tilde{r}-1$\\
   \hline
    RD/E-mode/lL & $(4\pi^2T_0^2B_0^{-2}/5)n_\pm(T_0,B_0)\sigma_{\rm T}r_0$& $0$ & $-5/4$ & $0$ & $11/4$ & $\tilde{r}^{3/2}-1$\\
   \hline
      MD/E-mode/lL & $(4\pi^2T_0^2B_0^{-2}/5)n_\pm(T_0,B_0)\sigma_{\rm T}r_0$& $-11/6$ & $3/2$ & $1/6$ & $5/2$&$\eta^{1/3}\tilde{r}-1$\\
   \hline
  \end{tabular}
\end{table*}

As the fireball expands, photons begin to escape from the fireball in the direction parallel or perpendicular to the magnetic field  (where the direction is measured in the comoving frame of the fluid).
At first, the fireball is optically thick and the thermal equilibrium is sustained.
In this case, the number density of the positron is that of the thermal equilibrium.\footnote{
Strictly speaking, the equilibrium value is not the same as that in Equation~(\ref{eq:number_T}) or (\ref{eq:number_BT}) because the chemical potential has a finite value due to the presence of baryons \citep[][Section~105]{Landau1980stat}.
However, we ignore this chemical potential for simplicity, and this approximation does not change the results significantly.}
The number density of baryon, $n_{\rm b}=\rho/m_{\rm p}$, where $m_{\rm p}$ is the proton mass, evolves as shown in Equation~(\ref{eq:sol_rho}), and the electron number density, $n_-$, is determined by the condition of charge neutrality,
\begin{equation}
  n_-=n_{\rm b}+n_+.
\end{equation}
For simplicity, we assume that all baryons are composed of protons.

First, we outline the case where photons escape in the direction parallel to the magnetic field (right part of Figure~\ref{fig:overview}).
As the fireball expands, the number density of the lepton decreases.
When the optical depth in the direction parallel to the magnetic field,
\begin{equation}
\tau_{\parallel}=(n_++n_-)\sigma(T,B) \frac{r}{\Gamma},
\end{equation}
becomes lower than unity, the fireball becomes optically thin, and the photons escape from the fluid in the direction longitudinal to the magnetic field line.

Next, we outline the case where photons escape in the direction perpendicular to the magnetic field (left part of Figure~\ref{fig:overview}).
As the number density of the lepton drops, the diffusion time of the photons in the direction perpendicular to the flux tube,  
\begin{equation}
t_{\rm diff}= \frac{\ell \tau_\perp}{c},
\end{equation}
becomes short, where the diffusion time is measured in the comoving frame of the fluid and 
\begin{equation}
    \tau_\perp=(n_++n_-)\sigma(T,B) \ell.\label{eq:tau_perp}
\end{equation}
When the diffusion timescale becomes shorter than the dynamical timescale,
\begin{equation}
    t_{\rm dyn}=\frac{r}{c\Gamma},\label{eq:t_dyn}
\end{equation}
the photons diffuse out laterally from the flux tube.
Afterwords, if the timescale for the pair creation is shorter than the dynamical timescale, the high-energy tail of the diffusing photons creates $e^{\pm}$ pairs outside the initial flux tube.
As a result, the leptonic fireball expands laterally.
On the contrary, if the timescale for the pair creation is longer than the dynamical timescale, the pair creation in the surrounding magnetic field lines is weak.
As a result, the photons just escape from the fireball and the baryons and leptons remain confined in the initial flux tube (left part of Figure~\ref{fig:overview} without the pair fireball outside the initial flux tube (yellow region)).
Because the lateral diffusion occurs only when the causal region is larger than the lateral size of the fireball,\footnote{
Diffusion occurs if both $t_{\rm diff}<t_{\rm dyn}$ and $\tau_\parallel>1$ are satisfied.
The former means $(n_++n_-)\sigma(T,B) \ell^2 <r/\Gamma$ and using the latter inequality, $(n_++n_-)\sigma(T,B)r/\Gamma>1$, we obtain $\ell<(r/\Gamma)$.
Thus, the lateral size of the fireball, $\ell$, is smaller than the causal region, $r/\Gamma$ when the diffusion occurs.}
the photons diffuse out from the fireball uniformly.
This diffusion occurs if the fireball is confined in a narrow flux tube ($\theta\lesssim10^{-2}$, see Section~\ref{fluid}).
Thus, in the context of the classical fireball for gamma-ray bursts, this lateral diffusion does not occur and have not discussed so far.

$\tau_\parallel$ and $t_{\rm diff}/t_{\rm dyn}$ determine where the photons escape.
Because the cross section for the lepton is three orders of magnitude larger than that for the baryon, the opacity is dominated by the leptons.
$\tau_\parallel$ and $t_{\rm diff}/t_{\rm dyn}$ are expressed as
\begin{align}
 \tau_\parallel&=(n_++n_-)\sigma(T,B)\frac{r}{\Gamma},\label{eq:tau_def}\\
 \frac{t_{\rm diff}}{t_{\rm dyn}}&=(n_++n_-)\sigma(T,B)\ell^2\frac{\Gamma}{r}\label{eq:diff_def}.
\end{align}
If the number density of the electrons associated with the baryon (baryonic electron component) is higher than that of the pair plasma, the number density of the lepton is determined by the baryonic electron component as
\begin{equation}
    n_++n_-\simeq n_{\rm b}.
\end{equation}
If the fireball contains few baryons and the number density of the baryonic electron component is lower than that of the pair plasma, the number density is determined by the pair plasma as
\begin{equation}
    n_++n_-\simeq2n_+.
\end{equation}

If the number density of the lepton is determined by the baryonic electron component, $n_{\rm b}=\rho/m_{\rm p}$, in Equation~(\ref{eq:sol_rho}), there are four cases depending on whether radiation-dominated or matter-dominated, and whether the suppression of the cross section is effective or not.
The analytical expressions of $\tau_{\rm \parallel}$ and $t_{\rm diff}/t_{\rm dyn}$ are
\begin{eqnarray}
 \tau_\parallel&=&\tau_{\rm b0}\eta^{\gamma}\tilde{r}^\delta,\label{eq:tau_b}\\
 \frac{t_{\rm diff}}{t_{\rm dyn}}&=&\tau_{\rm b0}\theta_0^2\eta^{\epsilon}\tilde{r}^\zeta,\label{eq:diff_b}
\end{eqnarray}
where $\tau_{\rm b0}$ and $\gamma$--$\zeta$ are given in Table~\ref{tab:index_baryon}.
When $\tilde{r}_{\rm E},\tilde{r}_{\rm S}>\tilde{r}$ (RD/E-mode in Table~\ref{tab:index_baryon}), $\tau_\parallel$ and $t_{\rm diff}/t_{\rm dyn}$ increase as the radius increases.
This is because the larger the radius, the weaker the suppression of the cross section for E-mode photons.
The increase in the cross section exceeds the decrease in the number density, and therefore $\tau_\parallel$ and $t_{\rm diff}/t_{\rm dyn}$ increase as the radius increases.

If the number density of the leptons is determined by the pair plasma, one new case arises in addition to the cases for the baryonic electron component.
This is because the thermal equilibrium number density also depends on the magnetic field (see Sec.~\ref{sec:number} and $r_{\rm B}$ in Table~\ref{tab:radbar_summary}).
Taking this into account, the analytical expressions of $\tau_{\rm \parallel}$ and $t_{\rm diff}/t_{\rm dyn}$ are
\begin{eqnarray}
 \tau_\parallel&=&\tau_{\rm \pm0}\eta^{\gamma^\prime}\tilde{r}^{\delta^\prime}\exp \left(-T_0^{-1}A\right),\label{eq:tau_p}\\
 \frac{t_{\rm diff}}{t_{\rm dyn}}&=&\tau_{\rm \pm0}\theta_0^2\eta^{\epsilon^\prime}\tilde{r}^{\zeta^\prime}\exp\left(-T_0^{-1}A\right),\label{eq:diff_p}
\end{eqnarray}
where $\tau_{\rm \pm0}$, $\gamma^\prime$--$\zeta^\prime$, and $A$ are given in Table~\ref{tab:index_pair}.
For pair plasma, the number density depends exponentially on the temperature, and the photospheric radius and lateral-diffusion radius are mainly determined by $A$.

The larger one of the photospheric radii for the pair plasma and baryonic electron component is the physical photospheric radius.
Namely, by solving the equation $\tau_\parallel=1$ for both cases of Equations~(\ref{eq:tau_b}) and (\ref{eq:tau_p}), we obtain the  photospheric radii for the baryonic electron component, $\tilde{r}_{\rm \parallel,b}$, and pair component, $\tilde{r}_{\rm \parallel,\pm}$, and choose the larger one.
The same procedure is applied to the lateral-diffusion radius.
The lateral-diffusion radius is the larger one of the solutions of $t_{\rm diff}/t_{\rm dyn}=1$ for the cases of Equation~(\ref{eq:diff_b}), $\tilde{r}_{\rm diff,b}$, and for the case of Equation~(\ref{eq:diff_p}), $\tilde{r}_{\rm diff,\pm}$.
Given the lateral-diffusion radius and the photospheric radius calculated in this way, the smaller one determines the radius where the photons escape.

For the baryonic electron component, the photospheric radius and the lateral-diffusion radius are
\begin{align}
\tilde{r}_{\rm \parallel,b}&=
\begin{dcases}
\tau_{\rm b O 0}^{1 / 5} \eta^{-1 / 5} &({\rm RD/O-mode})\\
\tau_{\rm b O 0}^{1 / 2} \eta^{-3 / 2}  &({\rm MD/O-mode})\\
\tau_{\rm b E 0}^{1 / 2} \eta^{-1 / 2} &({\rm RD/E-mode}) \\
\tau_{\rm b E 0}^{-1 / 2} \eta^{11 / 6} &({\rm MD/E-mode})
\end{dcases},\label{eq:radius_pb}\\
\tilde{r}_{\rm diff,b}&=
\begin{dcases}
\tau_{\rm b O 0}\theta_{0}^{2} \eta^{-1}  &({\rm RD/O-mode})\\
\tau_{\rm b O 0}\theta_{0}^{2} \eta^{-1}  &({\rm MD/O-mode}),\\
\tau_{\rm b E 0}^{-1 / 2} \theta_{0}^{-1} \eta^{1 / 2} &({\rm RD/E-mode}) \\
\tau_{\rm b E 0}^{-1 / 3} \theta_{0}^{-2 / 3} \eta^{5 / 9} &({\rm MD/E-mode})
\end{dcases}\label{eq:radius_db}
\end{align}
where $\tau_{\rm b O 0}=e_{0} \sigma_{\rm T} r_{0}/(m_{\rm p}c^2)$ and $\tau_{\rm b E 0} =(4 \pi^{2}T_0^2B_0^{-2}/5)e_{0} \sigma_{\rm T} r_{0}/(m_{\rm p}c^2)$, both of which do not depend on $\eta$.
These radii are power functions of $\eta$.
The lateral-diffusion radius for the pair plasma is mainly determined by the exponential part of the number density, $A$, and it is independent of $\eta$ in the radiation-dominated phase.

For the parameters that we are interested in ($B_0\sim{\mathcal O}(1)$ and $T_0\sim{\mathcal O}(0.1)$), in many cases, $\tilde{r}_{\rm diff,\pm}<\tilde{r}_{\rm \parallel,\pm}$ is satisfied, and we mainly consider this case.
The case of $\tilde{r}_{\rm diff,\pm}\geq\tilde{r}_{\rm \parallel,\pm}$ is discussed in Appendix.~\ref{sec:thin}.

\subsection{Radiative Acceleration}\label{sec:radacc}
After the fireball becomes optically thin, if the photospheric luminosity is higher than the kinetic luminosities, the pairs and the baryons can be accelerated by the escaping photons \citep{MesLag1993,GriWas1998,NakPir2005}.
The acceleration continues as long as the work done by the radiation
during the dynamical time in the comoving frame is higher than the rest mass energy of a particle.
In the case of expanding fireball of a magnetar, the radiative acceleration via cyclotron resonant scattering can be a dominant scattering process \citep{MitPav1982}.
Thus, we evaluate the final Lorenz factor realized by the radiative acceleration via resonant scattering.

The cross section for the resonant scattering averaged by polarization is \citep[e.g.,][]{CanLod1971,Ven1979}
\begin{equation}
  \sigma_{\rm res}=\frac{\pi^2 e^2}{m_{\rm e} c}\left(1+\cos^2\theta_{\rm kB}\right)\delta\left(\omega-\omega_c\right),
\end{equation}
where $\omega$ is the angular frequency, $\omega_{\rm c}=e B_{\rm Q} B/m_{\rm e}c$ is cyclotron angular frequency, $\theta_{\rm kB}$ is the angle between the magnetic field line and direction of the photon ray, and $\delta(x)$ is the delta function.
$\omega$, $\omega_{\rm c}$, and $\theta_{\rm kB}$ are measured in the electron rest frame.
As long as the electrons moves along the magnetic field, the magnetic field in the electron rest frame equals that in the lab frame, in which the magnetar is at rest.
During the acceleration, the electron is almost at rest in the frame where the photon is isotropic, but at the late phase of the acceleration, the electron is delayed behind the frame.
For this electron, the photon field is not isotropic and the electron is mainly accelerated by the photons with $\theta_{\rm kB}\simeq 0$.
Therefore, we set $\theta_{\rm kB}\simeq0$ because the photons with $\theta_{\rm kB}\simeq 0$ mostly contribute to the acceleration.
We also assume that the fireball expand almost radially, i.e., the fireball is not formed far from the magnetic pole.
The case that the fireball is formed near the equatorial plane,
$\vartheta\simeq \pi/2$, is discussed later.

Taking resonant scattering into account, the work done by the radiation in the comoving frame of a particle during the dynamical time equals \citep[e.g.,][]{ThoLyu2002}
\begin{align}
  \int d\omega\,\sigma_{\rm res}\frac{L_{\rm \omega,iso}}{4\pi r^2c \Gamma^2}\times\frac{r}{\Gamma}&=\frac{\pi e^2 \left(L_{\omega, \rm iso}\right)_{\omega=\omega_{\rm c}}}{2m_{\rm e}c^2r\Gamma^3}
\end{align}
where $L_{\omega, \rm iso}$ is the isotropic specific photospheric luminosity (i.e., $dL/d\omega$).
For the particle to be accelerated to relativistic speed, this work must be higher than the rest mass energy of the particle,
\begin{equation}
  \frac{\pi e^2 \left(L_{\omega, \rm iso}\right)_{\omega=\omega_{\rm c}}}{2m_{\rm e}c^2r\Gamma^3}>\bar{m}c^2 ,
\label{eq:CD_cond}
\end{equation}
where
\begin{equation}
\bar{m}\simeq\frac{m_{\rm p}n_{\rm b}+m_{\rm e}(2n_+)}{n_{\rm b}+2n_+},\label{eq:barm}
\end{equation}
is the mean rest mass of the outflow \citep{NakPir2005}.
Particles are accelerated such that the particles see the radiation field being almost isotropic, otherwise the radiation field is anisotropic and drags the particles.
The Lorentz factor of this isotropic-radiation frame is asymptotically \citep[e.g.,][]{GriWas1998},
\begin{equation}
 \Gamma\simeq\Gamma_{\parallel}\left(\frac{r}{r_{\rm \parallel}}\right),
 \label{eq:CD_evol}
\end{equation}
where $\Gamma_{\parallel}$ is the Lorentz factor at the radius where the fireball becomes optically thin, 
$r_{\rm \parallel}=\max(r_{\rm \parallel,b},r_{\rm \parallel,\pm})$,
we assume that the radiation comes from a spherical cap whose half-opening angle equals $1/\Gamma_{\parallel}$,\footnote{
Even if we assume that the radiation comes from a spherical cap whose half-opening angle equals that of the flux tube, the results do not change by an order of magnitude.
} and ignore order-of-unity
coefficient.
The acceleration ends where inequality~(\ref{eq:CD_cond}) fails.
Assuming the black body spectrum,
\begin{equation}
  L_{\rm \omega,iso}=L_{\mathrm{ph,iso}}\frac{15}{\pi^4}\left(\frac{\hbar \omega}{m_{\rm e}c^2 T}\right)^{3} \frac{\hbar}{m_{\rm e}c^2 T} \frac{1}{\exp\left(\hbar\omega/(m_{\rm e}c^2T)\right)-1},
\end{equation}
 where $L_{\rm ph,iso}$ is the isotropic luminosity of the radiation, with Equations~(\ref{eq:Br}), (\ref{eq:CD_cond}), and (\ref{eq:CD_evol}), and using $T\propto r^{-1}$ in the optically thin regime \citep{LiSar2008}, the final Lorentz factor is
\begin{align}
  \Gamma_{\rm RA}(r_{\rm \parallel})&\simeq\left(\frac{45}{4\pi^{3} \alpha}\right)^{1 / 7}\left(\frac{\sigma_{\rm T}L_{\rm ph,iso}}{4\pi \bar{m}c^3 r_{\rm \parallel}}\right)^{1 / 7}\nn\\
 &\quad \times\Gamma(r_{\rm \parallel})^{4 / 7} B(r_{\rm \parallel})^{2 / 7} T(r_{\rm \parallel})^{-3 /7} \label{eq:Gamma_res}\\
 &=:f_{\rm RA} \tilde{r}_{\rm \parallel}^{1/2},\nn
\end{align}
where we used $\exp(B/T)-1\simeq B/T$ for $B\ll T$, which is valid for the parameters currently being considered, used Equations~(\ref{eq:er_rad})--(\ref{eq:rho_rad}) at the last line, and
\begin{align}
    f_{\rm RA}&=\left(\frac{45}{4\pi^3\alpha}\right)^{1/7}\left(\frac{\sigma_{\rm T}L_{\rm ph,iso}}{4\pi \bar{m}c^3 r_{0}}\right)^{1/7}B_0^{2/7}T_0^{-3/7}\label{eq:fra}\\
   &\ \sim 26 \,L_{\rm ph,iso,41}^{1/7}\left(\frac{\bar{m}}{m_{\rm e}}\right)^{-1/7}B^{2/7}_{0,14}T_{0,\rm 100\,keV}^{-3/7},\nn
\end{align}
where $T_{0,\rm 100\,keV}=m_{\rm e}c^2T_0/100\,{\rm keV}$.
If the Lorentz factor in Equation~(\ref{eq:Gamma_res}) is higher than $\eta$, it means that the fireball becomes matter-dominated before the Lorentz factor reaches $\Gamma_{\rm RA}(r_{\rm \parallel})$.
In this case, the radiative acceleration ends at the radius where the Lorentz factor reaches $\eta$.
The electrons can be resonantly scattered not only by the photons near the peak of the blackbody spectrum but by the photons in the Rayleigh-Jeans regime.
Because the cross section for the resonant scattering is large, the radiative force, which is proportional to the product of the cross section and the radiative flux, can be strong enough to accelerate the outflow even if the photons near the peak of the balackbody spectrum do not contribute to the rariative force.

The observed X-ray spectrum have soft low-energy spectral index, and the radiative acceleration with this spectrum may realize higher Lorentz factor than the black body spectrum.
This is because there are more soft photons than with the black body spectrum, and the soft photons resonantly scatters with the electrons at the outer region.

If the fireball occurs in non-polar directions (at large $\vartheta$), the final Lorentz factor would be determined by $\Theta_{\rm kB}$, which is the angle between the magnetic field line and the direction of the photon ray in the lab frame.
A particle is accelerated so that the radiation force is directed perpendicularly to the magnetic field line in the comoving frame of the particle \citep{ThoLyu2002}.
This condition gives $\Gamma=1/\sin\Theta_{\rm kB}$.
Also, if the flux tube spread so wide that the particles along the tube leave the beamed radiated region during the acceleration, the radiative acceleration also stops.

For the parameters shown in Figure~\ref{fig:typical_r}, the inner part of the flux tube satisfying $\ell\lesssim3\times10^3\,{\rm cm}$ at the surface of the magnetar is accelerated up to $\Gamma_{RA}(r_\parallel)$ while the final Lorentz factor of the other part ($3\times10^{3}\,{\rm cm}\lesssim \ell<1\times10^{4}\,{\rm cm}$ at the surface of the magnetar) is determined by the angle between the magnetic field and the direction of the photon ray in the lab frame.
For the parameters in Figure~\ref{fig:200428}, the outflow in the flux tube with $\ell\lesssim7\times10^3\,{\rm cm}$ at the surface of the magnetar is accelerated up to $\Gamma_{RA}(r_\parallel)$, that is, all of the fireball is accelerated up to $\Gamma_{RA}(r_\parallel)$.
These values are derived assuming that the magnetic field is dipolar and the photon ray is radial.
If the flux tube points to the radial direction more than the dipole field, 
$\Gamma_{RA}(r_\parallel)$ would be realized more easily.

\section{Outflow and luminosity}\label{sec:cases}
\begin{figure}
\includegraphics[width=\linewidth]{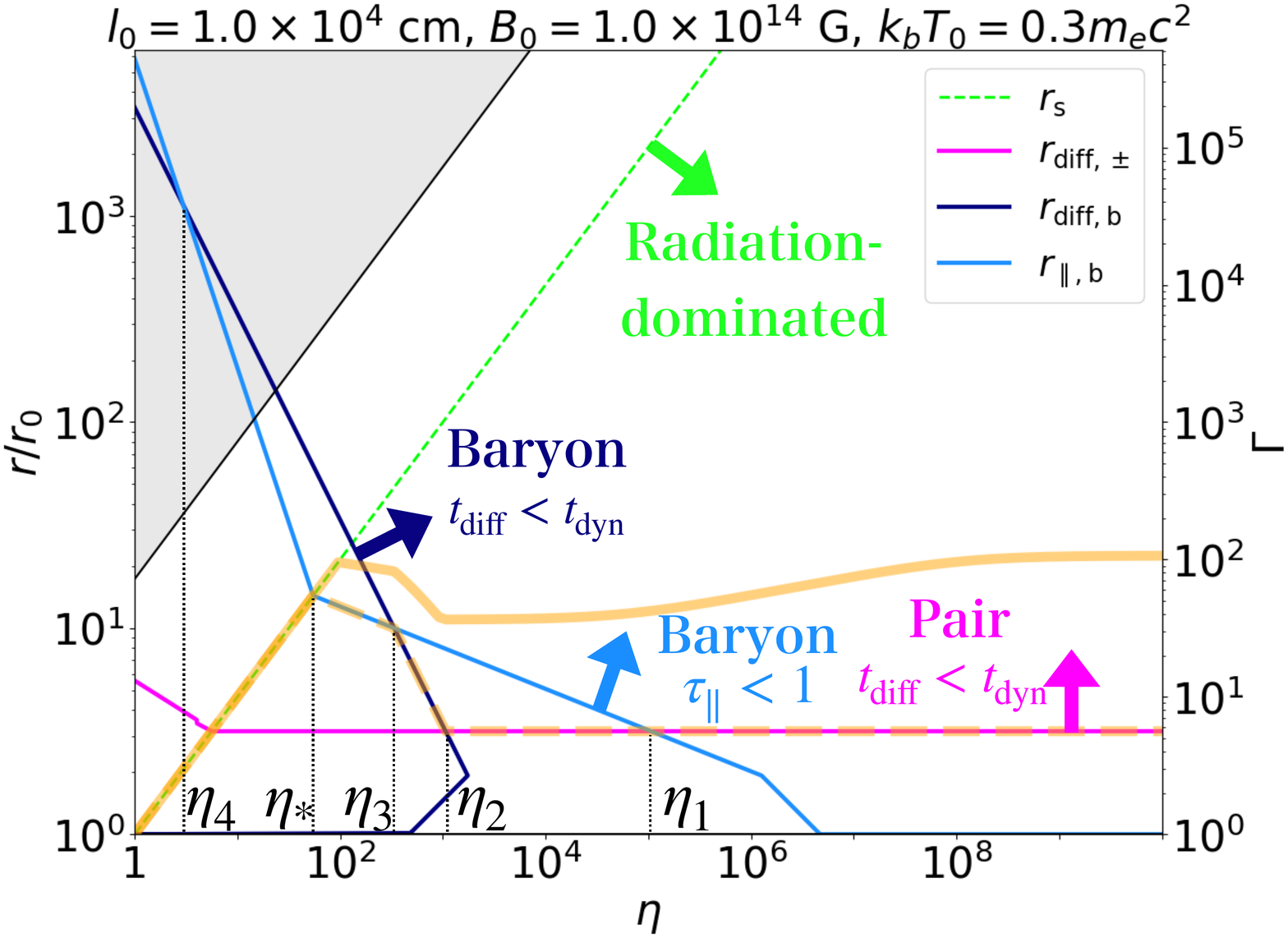}
\caption{
  $\eta$-dependence of the saturation radius, $\tilde{r}_{\rm S}$ (green dashed line, Equation~(\ref{eq:r_s})), the lateral-diffusion radius for pair component, $\tilde{r}_{\rm diff,\pm}$ (pink, see Equation~(\ref{eq:diff_p})), the lateral-diffusion radius for baryonic electron component $\tilde{r}_{\rm diff, b}$ (navy, Equation~(\ref{eq:radius_db})), the photospheric radius for baryonic electron component $\tilde{r}_{\rm \parallel, b}$ (light blue, Equation~(\ref{eq:radius_pb}), scale on the left axis), and the final Lorentz factor (yellow, scale on the right axis).
    The gray shaded region is above the Alfv\'{e}n radius where the assumption of expansion along the flux tube is violated.
The yellow solid and dashed lines shows the final Lorentz factor of the fireball with and without radiative acceleration beyond the photospheric radius, respectively (see Equations~(\ref{eq:sol_gamma}) and (\ref{eq:Gamma_res})).
  For the yellow solid line, the left axis does not show the radius where the final Lorentz factor is realized.
  This is because the radial dependence of the Lorentz factor in the optically thick region (Equation~(\ref{eq:gamma_rad})) is different from that in the optically thin region (Equation~(\ref{eq:CD_evol})).
  The photospheric radius for the pair plasma, $\tilde{r}_{\parallel,\rm p}$, is just above $\tilde{r}_{\rm diff,\pm}$ and we omit it for ease of viewing.
  In the region above the pink line and on the right side of the navy line, the diffusion time is shorter than the dynamical timescale.
  In the region above the pink line and sky blue line, the optical depth is lower than unity.
  Below the green dashed line, the fireball is radiation dominant.
$\eta_1$--$\eta_4$, and $\eta_*$ are also shown (Equations~(\ref{eq:eta1})--(\ref{eq:eta*})).
\label{fig:typical_r}}
\end{figure}

There are five different cases of the fireball evolution, more than that without the baryon loading.
Figure~\ref{fig:typical_r} shows $\eta$ dependence of each radius, $\tilde{r}_{\rm diff,\pm}$, $\tilde{r}_{\rm diff,b}$, and $\tilde{r}_{\rm \parallel,b}$.
We adopt $\ell_0=1\times10^4\,{\rm cm}$, $B_{\rm Q}B_0=1\times 10^{14}\,{\rm G}$, and $T_0=0.3$ (see Equation~(\ref{eq:normalize})).
All the five different cases of the evolution appear under these parameters.
The value of $\eta$ at the intersection of lines in $\eta$-$\tilde{r}$ plane, $\eta_1$--$\eta_4$, are defined as
\begin{eqnarray}
    \tilde{r}_{\rm diff,\pm}(\eta_1)&=&\tilde{r}_{\rm \parallel,b}(\eta_1),\label{eq:eta1}\\
    \tilde{r}_{\rm diff,\pm}(\eta_2)&=&\tilde{r}_{\rm diff,b}(\eta_2),\\
    \tilde{r}_{\rm diff,b}(\eta_3)&=&\tilde{r}_{\rm \parallel,b}(\eta_3)<\tilde{r}_{\rm S}(\eta_3),\\
    \tilde{r}_{\rm diff,b}(\eta_4)&=&\tilde{r}_{\rm \parallel,b}(\eta_4)>\tilde{r}_{\rm S}(\eta_4),\label{eq:eta4}
\end{eqnarray}
where $\tilde{r}_{\rm X}(\eta)$ (${\rm X}=({\rm diff,\pm}), \,({\rm diff,b}),\,({\rm \parallel,b}),$ and ${\rm S}$) represents the line of $\tilde{r}=\tilde{r}_{\rm X}(\eta)$ in $\eta$-$\tilde{r}$ plane (see Figure~\ref{fig:typical_r} and Equations~(\ref{eq:r_s}) and (\ref{eq:diff_p})--(\ref{eq:radius_db})).
We also define $\eta_*$ as the solution of the equations
\begin{equation}
  \tilde{r}_{\rm S}(\eta_*)=\min \left[\tilde{r}_{\rm diff,b}(\eta_*),\tilde{r}_{\rm \parallel,b}(\eta_*)\right],\label{eq:eta*}
\end{equation}
(see Equations~(\ref{eq:r_s}), (\ref{eq:radius_pb}), and (\ref{eq:radius_db})).

\begin{itemize}
\item[(1)] For $\eta_1<\eta$ (pair-diffusion case), the baryonic electron component does not contribute to the optical depth and the evolution is the same as the fireball without baryons.
The photons diffuse out from the initial flux tube laterally and we call this case ``pair-diffusion case''.
\item[(2)] For $\eta_2<\eta\leq\eta_1$ (pair-diffusion-baryon-photosphere case), the baryonic electron component does not contribute to the diffusion radius but does contribute to the photospheric radius.
In this case, the lateral-diffusion radius is determined by the pair plasma, i.e., $\max(r_{\rm diff,\pm},r_{\rm diff,b})=r_{\rm diff,\pm}$.
However, for $r>r_{\rm diff,\pm}$, the baryonic electrons in the initial flux tube make the outflow optically thick at the photospheric radius, i.e., $r_{\rm \parallel,b}>r_{\rm diff,\pm}$.
We call this case ``pair-diffusion-baryon-photosphere case''.
\item[(3)] For $\eta_3<\eta\leq\eta_2$ (baryon-diffusion case), the baryonic electron component contributes to both diffusion timescale and optical depth, and the lateral-diffusion radius is smaller than the photospheric radius, i.e., $r_{\rm \parallel,b}>r_{\rm diff,b}>r_{\rm diff,\pm}$
We call this case ``baryon-lateral case''.
\item[(4)] For $\eta_*<\eta\leq\eta_3$ (baryon-photosphere case), the baryonic electron component determines both diffusion timescale and optical depth, and the photospheric radius is smaller than the lateral-diffusion radius, i.e., $r_{\rm diff,b}>r_{\rm \parallel,b}>r_{\rm diff,\pm}$.
We call this case ``baryon-photosphere case''.
\item[(5)] For $\eta\leq\eta_*$ (baryon-dominant case), the fireball becomes matter-dominated phase at $\tilde{r}=\tilde{r}_{\rm S}$.
We call this case ``baryon-dominant case''.
In the baryon-dominant case, the photons escape from the fireball at smaller radius of $\tilde{r}=\tilde{r}_{\rm\parallel,b}$ ($\eta_4<\eta\leq\eta_*$) and $\tilde{r}=\tilde{r}_{\rm diff,b}$ ($\eta<\eta_4$).
\end{itemize}
In the following sub-sections, we consider these cases in detail.

In Figure \ref{fig:typical_r}, the Lorentz factor resulting from the radiative acceleration due to resonant scattering is shown by the yellow solid line.
This line corresponds to the case that the fireball occurs in the polar direction ($\vartheta\sim0$) and the flux tube does not spread too wide.
In the case that the photons escape from the fireball after it becomes diffusively thin ($\eta_3<\eta$), this acceleration occurs after the fireball becomes optically thin.
If the pair plasma dominates the optical depth ($\eta_1<\eta$), the photospheric radius is just above the lateral diffusion radius, and this acceleration occurs.
If the baryonic electron component dominates the optical depth and the photons escape from the fireball via lateral diffusion ($\eta_3<\eta<\eta_1$), the photospheric radius can be much larger than the lateral diffusion radius (see Figure~\ref{fig:typical_r}).
In this case, if the radius where acceleration stops, $r_{\rm \parallel}\Gamma_{\rm RA}/\Gamma_{\rm \parallel}$, is larger than the photopheric radius, $r_{\rm \parallel, b}$, the radiative acceleration occurs.
Otherwise, above the photospheric radius, the work done by the radiation pressure is not enough to accelerate the fluid.
For the parameters of Figure~\ref{fig:typical_r}, the condition for the radiative acceleration is satisfied.
If the radius where acceleration stops, $r_{\rm \parallel}\Gamma_{\rm RA}/\Gamma_{\rm \parallel}$, is near the photopheric radius, $r_{\rm \parallel, b}$, some of the photons from the newly created pair plasma fireball may not be able to contribute to the acceleration.
This is because they can be scattered by the particle in the optically thick fireball ($r<r_{\rm \parallel,b}$) before they reaches the particles in the optically thin region ($r\geq r_{\rm \parallel,b}$).
This effect may change $\Gamma_{\rm RA}$.

\subsection{Pair-diffusion case \texorpdfstring{($\eta_1<\eta$)}{}}\label{sec:pl}
If the fireball contains almost no baryons, the evolution is the same as that of a purely pair fireball along open magnetic field lines \citep{Iok2020}.
At the lateral-diffusion radius, photons begin to diffuse laterally from the initial flux tube, and the diffusing photons create electron-positron pairs out of the initial flux tube (see the left of Figure~\ref{fig:overview}).
The size of the fireball increases in the direction perpendicular to the magnetic field lines from $\ell(r_{\rm diff,\pm})$ to $\ell_{\rm diff}$, and the temperature decreases from $T(r_{\rm diff,\pm})$ to $T_{\rm diff}$.
Then, the number density of the pair plasma also decreases, finally the fireball becomes optically thin in the direction longitudinal to the magnetic field lines.
$T_{\rm diff}$ is determined by solving 
\begin{equation}
  2n_+\left(T_{\rm diff},B(r_{\rm diff,\pm})\right)\sigma\left(T_{\rm diff},B(r_{\rm diff,\pm})\right)\frac{r_{\rm diff,\pm}}{\Gamma(r_{\rm diff,\pm})}=1,
  \label{eq:thin_diff}
\end{equation}
where $r_{\rm diff,\pm}$ is the pink solid line and $\Gamma_(r_{\rm diff})$ is the yellow dashed line in Figure~\ref{fig:typical_r}.
If the total energy is conserved during the lateral expansion, the size of the fireball becomes 
\begin{equation}
    \ell_{\rm diff}=\ell(r_{\rm diff,\pm})\left(\frac{T(r_{\rm diff,\pm})}{T_{\rm diff}}\right)^2\label{eq:l_diff}.
\end{equation}
Equations~(\ref{eq:thin_diff}) and (\ref{eq:l_diff}) are evaluated at $r_{\rm diff,\pm}$.
This is because as the photons diffuse out, the temperature decreases, the number density of the pair plasma drops exponentially, and $t_{\rm diff}$ also drops exponentially while $t_{\rm dyn}$ does not change significantly.
The fact that $r_{\rm diff,\pm}$ is not significantly different from $r_{\rm \parallel,\pm}$ (a factor of 1.4 difference even for $\theta_0=10^{-3}$ and less for larger $\theta_0$) also justify this evaluation because $r_{\rm \parallel,\pm}$ (photospheric radius without lateral expansion) must be larger than the radius where Equation~(\ref{eq:thin_diff}) is satisfied (photospheric radius with lateral expansion).

Once the fireball becomes diffusively thin but optically thick for E-mode photons, O-mode photons also begin to escape.
This is because the mode exchange rate from O-mode to E-mode is comparable to the scattering rate of the E-mode photons, and the O-mode photons are converted to the E-mode photons in the optically thick (but diffusively thin) fireball.
For pair-diffusion-baryon-photosphere case (Section.~\ref{sec:pdbt}) and baryon-diffusion case (Section.~\ref{sec:baryon_diff}), this mode conversion would work because the difference between the diffusion radius and the photospheric radius is of the order of the dynamical radius except $\eta\sim \eta_1$ and $\eta\sim \eta_3$ (see Figure~\ref{fig:typical_r}).
For pair-diffusion case, the diffusion radius is close to the photospheric radius, and the sufficient conversion from the O-mode to the E-mode may not occur.
However, for the pair plasma, the diffusion radius and the photospheric radius for E-mode photons are very close to those of the O-mode photons.
The difference is less than $\sim 30\%$ between $B=10^{-1}$ case (no suppression) and $B=10^2$ case (extremely strong magnetic field), for the parameters discussed in Section.~\ref{sec:200428}.
We note that in the parameters of Figure~\ref{fig:typical_r}, $r_{\rm diff,\pm}$ is determined by the Thomson cross section ($r_{\rm diff,\pm}>r_{\rm E}$).
Therefore, around the diffusion (photospheric) radius for the E-mode photon, the O-mode photons also begin to escape from the fireball for all cases.

In the pair-diffusion case, the true photospheric luminosity (not isotropic luminosity), $L_{\rm ph}$, the true kinetic luminosity of the baryonic component, $L_{\rm kin}$, and the true kinetic luminosity of the pair component, $L_{\rm kin,\pm}$, in the lab frame are expressed as 
\begin{align}
  L_{\rm ph}&=\pi \ell_{\rm diff}^2caT_{\rm diff}^4\Gamma(r_{\rm diff,\pm})^2\label{eq:Lph_pp}\\
  &= L_0,\nn\\
  L_{\rm kin}&=\pi \ell(r_{\rm diff,\pm})^2\rho(r_{\rm diff,\pm})c^3\Gamma(r_{\rm diff,\pm})\Gamma_{\rm RA}(r_{\rm diff,\pm})\label{eq:Lkin_pp}\\
  &\simeq \begin{cases}
    L_{0} f_{\rm RA}\eta^{-1}\tilde{r}_{\mathrm{diff}, \pm}^{1 / 2} &(\Gamma_{\rm RA}(r_{\rm diff,\pm})<\eta)\nn\\
    L_0 &(\Gamma_{\rm RA}(r_{\rm diff,\pm})\geq\eta)
    \end{cases},\nn\\
  L_{\rm kin,\pm}&=\pi \ell_{\rm diff}^2 m_{\rm e}c^3\left(2n_{+}(T_{\rm diff})\right)\Gamma(r_{\rm diff,\pm})\Gamma_{\rm RA}(r_{\rm diff,\pm})\label{eq:Lkine_pp}\\
  &\simeq L_0f_{\rm RA}\left(\frac{\theta_0^{2}\pi m_{\rm e}c^3r_0}{L_0\sigma_{\rm T}}\right)\tilde{r}_{\rm diff,\pm}^{11/2}\left(\frac{\ell_{\rm diff}}{\ell(r_{\rm diff,\pm})}\right)^2\nn\\
  &\quad\times\left(\frac{\sigma_{\rm T}}{\sigma(T_{\rm diff},B(r_{\rm diff,\pm}))}\right)\nn\\
  &\sim 2\times10^{-5}L_0f_{\rm RA,1.5}\theta_{\rm 0, -2}^2L_{0,40}^{-1}r_{\rm diff,\pm,0.5}^{11/2}\nn\\
  &\quad \times\left(\frac{\ell_{\rm diff}}{\ell(r_{\rm diff,\pm})}\right)^2\left(\frac{\sigma_{\rm T}}{\sigma(T_{\rm diff},B(r_{\rm diff,\pm}))}\right),\nn
\end{align}
where we used Equations~(\ref{eq:sol_e})--(\ref{eq:sol_rho}), (\ref{eq:Gamma_res}), (\ref{eq:thin_diff}), and (\ref{eq:l_diff}), and set $r_{\rm parallel}=r_{\rm diff,\pm}$.
Because the number density flux is conserved (Equation~(\ref{eq:baryon_dyn}) for baryonic electron component), $\ell^2\rho\Gamma$ and $ \ell^2n_{\rm +}\Gamma$ in the kinetic luminosities are evaluated at $r_{\rm diff,\pm}$.
Because the total energy flux of photons are assumed to be conserved during the diffusion (Equation~(\ref{eq:l_diff})), $L_{\rm ph}$ equals that value without diffusion.
Because of the relativistic beaming, the photons are emitted in the half-opening angle of
\begin{equation}
  \theta_{\rm beam}\sim\Gamma(r_{\rm diff,\pm})^{-1},
\end{equation}
while the baryon, electron and positron flows along the flux tube.

\subsection{Pair-diffusion-baryon-photosphere case \texorpdfstring{($\eta_2<\eta\leq\eta_1$)}{}}\label{sec:pdbt}
In this case, the fireball in the initial flux tube is optically thick due to the baryonic electrons component just after lateral diffusion begins at $r_{\rm diff,\pm}$.
The lateral diffusion and pair creation outside the initial flux tube would occur as in the pair-diffusion case.
In the initial flux tube, the baryonic electron component is confined, and the photons there cannot escape in the direction parallel to the magnetic field.
However, the purely pair fireball created out of the initial flux tube becomes optically thin, where the photons can escape in the direction parallel to the field lines (see the left of Figure~\ref{fig:overview}).
Therefore, the photons first diffuse laterally from the initial fireball to the newly created pair plasma fireball, and then escape in the direction parallel to the field lines.
For simplicity, in evaluating the luminosities, we assume that the photons and pairs are emitted from the entire section of the fireball.
This approximation does not change the results significantly.\footnote{
For the parameters in Figure~\ref{fig:typical_r}, $\ell_{\rm diff}\sim1.6\ell(\tilde{r}_{\rm diff,\pm})$ and the observable area is $1-\left(1/1.6\right)^2\sim0.63$ of all region.
Thus, our approximation is valid within an accuracy of a factor of 2.\label{fn:l}}

In the initial flux tube, the optical depth of the pair annihilation for a positron is higher than unity.
This is because the baryonic electron component are able to annihilate with the positrons even if the density of the pair electrons drops.
However, we also ignore this effect in evaluating the kinetic luminosity of the pair-plasma outflow, $L_{\rm kin,\pm}$, because the number density of the pair plasma, which is two times the number density of the positron, is dominated by the newly created purely pair fireball, where the timescale of the pair annihilation is longer than the dynamical timescale.
In summary, the luminosities and the final Lorentz factor are the same as those of the pair-diffusion case (Equation~(\ref{eq:Lph_pp})--(\ref{eq:Lkine_pp})).

The final Lorentz factor of the baryons and pairs accelerated by the escaping radiation in Equation~(\ref{eq:Gamma_res}) may be different.
In the initial flux tube, the mean mass would be about $m_{\rm p}$ because positrons efficiently annihilate with the baryonic electron component.
In the newly created lateral fireball, the mean mass is $m_{\rm e}$ because there are no baryons.
Thus, the final Lorentz factor, which is proportional to $\bar{m}^{-1/7}$ (see Equation~(\ref{eq:Gamma_res})), is different in the initial flux tube and the newly created lateral fireball.
Because the kinetic luminosity is dominated by the baryonic component (see Figure~\ref{fig:typical_L}), we use $\bar{m}\sim m_{\rm p}$ to evaluate the final Lorentz factor in Figure~\ref{fig:typical_r}.
The final Lorentz factor and the kinetic luminosity of the lateral pairs component can be higher than the values in Figures~\ref{fig:typical_r} and \ref{fig:typical_L} by a factor of $(m_{\rm e}/m_{\rm p})^{-1/7}\sim 3$.

\subsection{Baryon-diffusion case \texorpdfstring{($\eta_3<\eta\leq\eta_2$)}{}}\label{sec:baryon_diff}
If the fireball contains more baryons, the diffusion timescale is determined by the baryonic electron component.
In many cases of the baryon-diffusion case, the diffusing photons cannot create electron-positron pairs out of the initial flux tube due to its low temperature.
Then, the photons escape by diffusing across the initial flux tube and propagate directly toward the observer.
The photospheric luminosity is determined by the amount of photons diffusing from the initial flux tube.
To evaluate the luminosity, we consider a portion of the initial flux tube at $r\sim r_{\rm diff,b}$ and approximate its shape as a cylinder.
The typical length scale of the flux tube is $c t_{\rm dyn}$ (Equation~(\ref{eq:t_dyn})), and the radius of the cross-section is $\ell(r_{\rm diff,b})$.
Thus, the area of the side is $2\pi \ell(r_{\rm diff,b})\times c t_{\rm dyn}$.
Since photons escape from this region by diffusion, their net speed is $c/\tau_{\parallel}$ (Equation~(\ref{eq:tau_perp})).
Therefore, the photospheric luminosity is 
\begin{eqnarray}
 L_{\rm ph}&=&2\pi \ell(r_{\rm diff,b})\times ct_{\rm dyn}\times aT^4(r_{\rm diff,b})\times\frac{c}{\tau_{\parallel}}\times\Gamma(r_{\rm diff,d})^2\nn\\
 &\simeq&2\pi \ell^2(r_{\rm diff,b})caT^4(r_{\rm diff,b})\Gamma(r_{\rm diff,d})^2,
\end{eqnarray}
where we have used ${\tau_{\parallel}}l/c\simeq r/(c\Gamma)$, which is valid for the lateral-diffusion radius.
This luminosity is twice as high as that of the previous cases (see Equation~(\ref{eq:Lkin_pp})), and we neglect this factor of two for simplicity.

In the baryon-diffusion case, the pair annihilation continues in the initial flux tube after the photons diffuse out because  baryonic electrons remains as targets.
The pair annihilation ceases at $r_{\rm \parallel,b}$, where the timescale of the pair annihilation for the positrons becomes longer than the dynamical timescale (Equation~(\ref{eq:t_dyn})).
During the pair annihilation, the number density of the positron follows the equation \citep[for details in the spherically symmetric case,][]{GriWas1998,YamKis2019},
\begin{equation}
    \frac{d}{dr}\left(n_+^*(r)r^2\Delta\Omega\right)\simeq-n_{\rm b}(r)\sigma_{\rm T} n_+^*(r)r^2\Delta\Omega,
\end{equation}
where $n_+^*(r)$ represents the comoving number density of the positron as a function of $r$, and we used the fact that the cross section for pair annihilation is approximated by $\sigma_{\rm T}/\beta_{\pm}$ for a small thermal velocity $\beta_\pm\ll1$.
Solving this equation in the radiation-dominated phase (see Equation~(\ref{eq:sol_rho})), the final number density of the positron is
\begin{eqnarray}
    n_+^*(r_{\rm \parallel, b}) &=&n_+^*(r_{\rm f})\left(\frac{\tilde{r}_{\rm f}}{\tilde{r}_{\parallel,b}}\right)^3\nonumber\\
    & &\times\exp\left[-\frac{2\rho_0\sigma_{\rm T}r_0}{7m_{\rm p}}\left(\tilde{r}_{\rm f}^{-7/2}-\tilde{r}_{\rm \parallel,b}^{-7/2}\right)\right]\label{eq:ne_b},
\end{eqnarray}
where $\tilde{r}_{\rm f}=\min\left[\tilde{r}_{\parallel,\pm},\tilde{r}_{\rm diff,b}\right]$.
At $\tilde{r}_{\rm f}$, the pair creation freeze out.
In many cases that we are interested in, $\tilde{r}_{\rm f}=\tilde{r}_{\parallel,\pm}\simeq \tilde{r}_{\rm diff,\pm}$ is satisfied.

To summarize, the luminosities and Lorentz factor are expressed as
\begin{align}
  L_{\rm ph}&\simeq \pi \ell^2(r_{\rm diff,b})caT^4(r_{\rm diff,b})\Gamma(r_{\rm diff,b})^2\label{eq:Lph_bd}\\
  &= L_0,\nn\\
  L_{\rm kin}&=\pi \ell^2(r_{\rm diff,b})\rho(r_{\rm diff,b})c^3\Gamma(r_{\rm diff,b})\Gamma_{\rm RA}(r_{\rm diff,b})\label{eq:Lkin_bd}\\
  &\simeq
  \begin{cases}
    L_{0} f_{\rm RA}\eta^{-1}\tilde{r}_{\mathrm{diff, b}}^{1 / 2} &(\Gamma_{\rm RA}(r_{\rm diff,b})<\eta)\nn\\
    L_0 &(\Gamma_{\rm RA}(r_{\rm diff,b})\geq\eta)
    \end{cases},\nn\\
  L_{\rm kin,\pm}&=\pi \ell^2(r_{\rm diff,b}) m_{\rm e}c^3\left(2n_{\rm +}^*(r_{\rm \parallel,b})\right)\Gamma(r_{\rm diff,b})\Gamma_{\rm RA}(r_{\rm diff,b})\label{eq:Lkine_bd}\\
  &\sim{\rm exponential\,\,decrease\,\,with \,\,decreasing \,\,}\eta,\nn  
\end{align}
where $n_{\rm +}^*(r_{\rm \parallel,b})$ is determined by Equation~(\ref{eq:ne_b}) and $f_{\rm RA}$ appears because of the radiative acceleration (see Equation~(\ref{eq:fra})).
Because $n_{\rm +}^*(r_{\rm \parallel,b})$ decreases exponentially as the radius increases and $r_{\rm \parallel,b}$ increases as $\eta$ decreases, $L_{\rm kin,\pm}$ drops exponentially as $\eta$ decreases.
$\eta$-dependence of $\tilde{r}_{\rm diff,b}$ is in Equation~(\ref{eq:radius_db}), and $\eta$-dependence of $L_{\rm kin}$ is also obtained using Equations~(\ref{eq:radius_db}) and (\ref{eq:Lkin_bd}).
\begin{align}
L_{\rm kin}\propto 
\begin{cases}
\eta^{-3/2} &({\rm RD/O-mode})\\
\eta^{-3/2} &({\rm MD/O-mode})\\
\eta^{-3/4} &({\rm RD/E-mode})\\
\eta^{-13/18} &({\rm MD/E-mode})\\
\end{cases}
\end{align}

\subsection{Baryon-photosphere case \texorpdfstring{($\eta_*<\eta\leq\eta_3$)}{}}\label{sec:blc}
We consider the case where the fireball contains so many baryons that the photons cannot escape from the fireball until it expands significantly.
In this case, the initial flux tube becomes laterally wide
(large $\ell(r)$) maintaining $t_{\rm diff}>t_{\rm dyn}$,
and $\tau_{\rm \parallel}<1$ is realized before $t_{\rm diff}$ becomes shorter than $t_{\rm dyn}$.
The photons, pair plasma, and baryonic components flow along the initial flux tube without diffusion up to the photospheric radius determined by the baryonic electron component, $\tilde{r}_{\rm \parallel,b}$ (Equation~(\ref{eq:radius_pb})).
The number density of the positrons decays following Equation~(\ref{eq:ne_b}) from $\tilde{r}_{\parallel,\pm}$ to $\tilde{r}_{\rm \parallel,b}$.
Thus, the luminosity is expressed as
\begin{align}
  L_{\rm ph}&=\pi \ell^2(r_{\rm \parallel,b})caT^4(r_{\rm\parallel,b})\Gamma(r_{\rm \parallel,b})^2\label{eq:Lph_bt}\\
  &= L_0,\nn\\
  L_{\rm kin}&=\pi \ell^2(r_{\rm \parallel,b})\rho(r_{\rm \parallel,b})c^3\Gamma(r_{\rm \parallel,b})\Gamma_{\rm RA}(r_{\rm \parallel,b})\label{eq:Lkin_bt}\\
  &\simeq  \begin{cases}
    L_{0}f_{\rm RA} \eta^{-1} \tilde{r}_{\rm\parallel, b}^{1 / 2} &(\Gamma_{\rm RA}(r_{\rm \parallel,b})<\eta)\nn\\
    L_0 &(\Gamma_{\rm RA}(r_{\rm \parallel,b})\geq\eta)
    \end{cases},\\
  L_{\rm kin,\pm}&=\pi \ell^2(r_{\rm \parallel,b}) m_{\rm e}c^3\left(2n_{\rm +}^*(r_{\rm \parallel,b})\right)\Gamma(r_{\rm diff,b})\Gamma_{\rm RA}(r_{\rm \parallel,b})\label{eq:Lkine_bt}\\
  &\sim {\rm exponential\,\,decrease\,\,with\,\,decreasing\,\,}\eta.\nn  
\end{align}
$f_{\rm RA}$ appears because of the radiative acceleration (see Equation~(\ref{eq:fra}))
Using $\eta$-dependence of $\tilde{r}_{\rm \parallel,b}$ in Equation~(\ref{eq:radius_pb}), that of $L_{\rm kin}$ is also obtained.
\begin{align}
L_{\rm kin}\propto 
\begin{cases}
\eta^{-11/10} &({\rm RD/O-mode})\\
\eta^{-7/4} &({\rm MD/O-mode})\\
\eta^{-5/4} &({\rm RD/E-mode})\\
\eta^{-1/12} &({\rm MD/E-mode})\\
\end{cases}
\end{align}

\subsection{Baryon-dominant case \texorpdfstring{($\eta\leq\eta_*$)}{}}
Finally, we consider the case where the fireball becomes matter dominated.
In this case, the acceleration of the fireball ends at $\tilde{r}_{\rm S}$.
At $\tilde{r}\simeq\tilde{r}_{\rm S}$, the fireball enters a coasting phase, and only the pair annihilation occurs.
Radiative acceleration (Section~\ref{sec:radacc}) does not occur in this case because the photospheric luminosity is lower than the kinetic luminosity of the baryonic component.
The luminosities are expressed as
\begin{align}
  L_{\rm ph}&=\pi \ell^2(r_{\rm b})caT^4(r_{\rm b})\Gamma(r_{\rm b})^2\label{eq:Lph_bm}\\
  &=\begin{cases}
  L_0\eta^{2/3}\tilde{r}_{\rm\parallel,b}^{-1}&(\eta_4<\eta\leq\eta_*)\\
  L_0\eta^{2/3}\tilde{r}_{\rm diff,b}^{-1}&(\eta\leq\eta_4)
  \end{cases},\nn\\
  L_{\rm kin}&=\pi \ell^2(r_{\rm b})\rho(r_{\rm b})c^3\Gamma(r_{\rm b})^2\label{eq:Lkin_bm}\\
  &=L_0,\nn\\
  L_{\rm kin,\pm}&=\pi \ell^2(r_{\rm b}) m_{\rm e}c^3\left(2n_{\rm +}^*(r_{\rm \parallel,b})\right)\Gamma(r_{\rm b})^2\label{eq:Lkine_bm}\\
  &\sim {\rm exponential\,\,decrease\,\,with\,\,decreasing}\,\,\eta,\nn  
\end{align}
where $r_{\rm b}=\min\left(r_{\rm \parallel,b},r_{\rm diff,b}\right)$.

\subsection{Summary of the Luminosity}\label{sec:summary_luminosity}
\begin{figure}
\includegraphics[width=\linewidth]{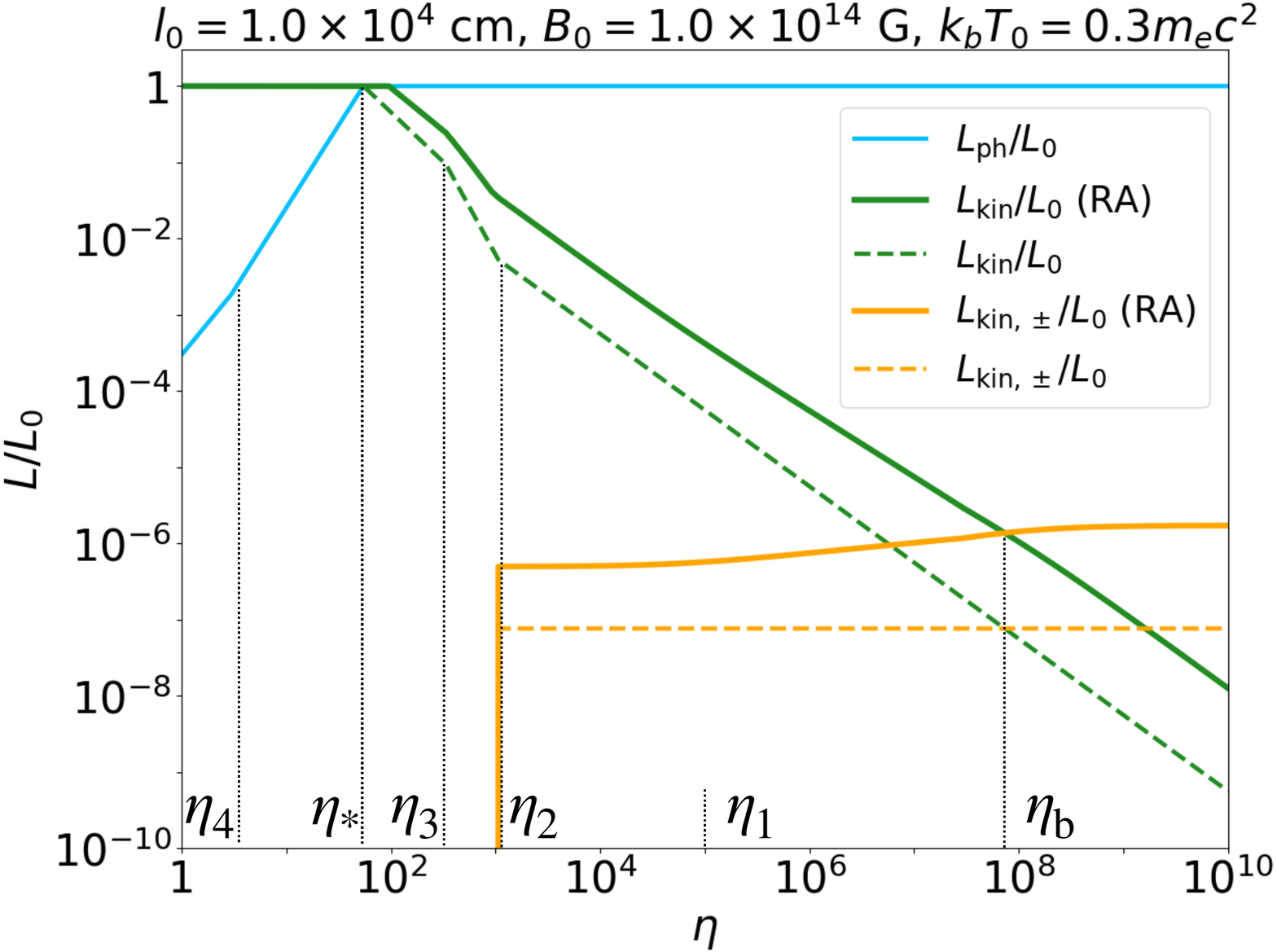}
\caption{
  Photospheric luminosity (light blue), kinetic luminosity of the baryonic component (green), and that of the pair component (orange; see Equations~(\ref{eq:Lph_pp})--(\ref{eq:Lkine_pp}), and (\ref{eq:Lph_bd})--(\ref{eq:Lkine_bm})).
  Each luminosity is normalized by the initial luminosity, $L_0=\pi \ell_0^2e_0c$.
  These luminosities are not isotropic luminosities but true luminosities.
  RA means radiative acceleration beyond the photospheric radius and the solid line represents the kinetic luminosity with radiative acceleration, and the dashed line represents that luminosity without radiative acceleration.
\label{fig:typical_L}}
\end{figure}

Figure~\ref{fig:typical_L} shows $\eta$-dependence of these (true) luminosities normalized by the injected luminosity,
\begin{equation}
    L_0=\pi \ell_0^2 e_0c.
\end{equation} 
In the pair-diffusion case and the pair-diffusion-baryon-photosphere case ($\eta_2<\eta$), almost all initial luminosity is emitted as the photospheric emission ($L_{\rm ph}=L_0$).
As $\eta$ decreases, the kinetic luminosity of the baryonic component increases because the amount of the baryonic component increases (green solid line).
For $\eta\lesssim \eta_{\rm b}$ (see Equation~(\ref{eq:eta_crit}) later), the baryon contributes inertia ($\bar{m}$ becomes higher than $m_{\rm e} $, see Equation~(\ref{eq:barm})) and the final Lorentz factor after the radiative acceleration beyond the photospheric radius decreases (Equation~(\ref{eq:Gamma_res})).
Thus, the kinetic luminosity of pair component also decreases (yellow solid line).
In the baryon-diffusion case ($\eta_3<\eta\leq\eta_2$), the slope of the kinetic luminosity of the baryonic component becomes steeper.
This is because the diffusive radius increases and the final Lorentz factor also increases as $\eta$ decreases.
In this region, the kinetic luminosity of the pair plasma drops exponentially due to the pair annihilation (Equation~(\ref{eq:ne_b})).
In the baryon-dominant case ($\eta<\eta_*$), the photospheric luminosity decreases as $\eta$ decreases.
This is because the initial radiation energy is converted into the kinetic energy of the baryonic component, as in the case of the spherically symmetric.

As $\eta$ decreases, at $\eta_{\rm b}$, the kinetic luminosity of the baryonic component exceeds that of the pair plasma.
Solving the equation $L_{\rm kin}=L_{\rm kin,\pm}$ for $\eta$, we obtain $\eta_{\rm b}$, below which the kinetic luminosity of the pair plasma is lower than that of the baryonic component.
Using Equations~(\ref{eq:Lkin_pp}) and (\ref{eq:Lkine_pp}), $\eta_{\rm b}$ is
\begin{align}
  \eta_{\rm b}&=\left(\frac{\theta_0^{2}\pi m_{\rm e}c^3r_0}{L_0\sigma_{\rm T}}\right)^{-1}\tilde{r}_{\rm diff,\pm}^{-5}\nn\\
  &\quad\times\left(\frac{\ell_{\rm diff}}{\ell(r_{\rm diff,\pm})}\right)^{-2}\left(\frac{\sigma_{\rm T}}{\sigma(T_{\rm diff},B(r_{\rm diff,\pm}))}\right)^{-1}  \label{eq:eta_crit}\\
  &\sim 3\times10^6\,L_{0,40}\theta_{0,-2}^{-2}\tilde{r}_{\rm diff,\pm,0.5}^{-5}\nn\\
  &\quad \times\left(\frac{\ell_{\rm diff}}{\ell(r_{\rm diff,\pm})}\right)^{-2}\left(\frac{\sigma_{\rm T}}{\sigma(T_{\rm diff},B(r_{\rm diff,\pm}))}\right)^{-1}.\nn
\end{align}
Below $\eta_{\rm b}$, the baryons also contributes the inertia.

\subsection{Observed Temperature}
\begin{figure}
\includegraphics[width=\linewidth]{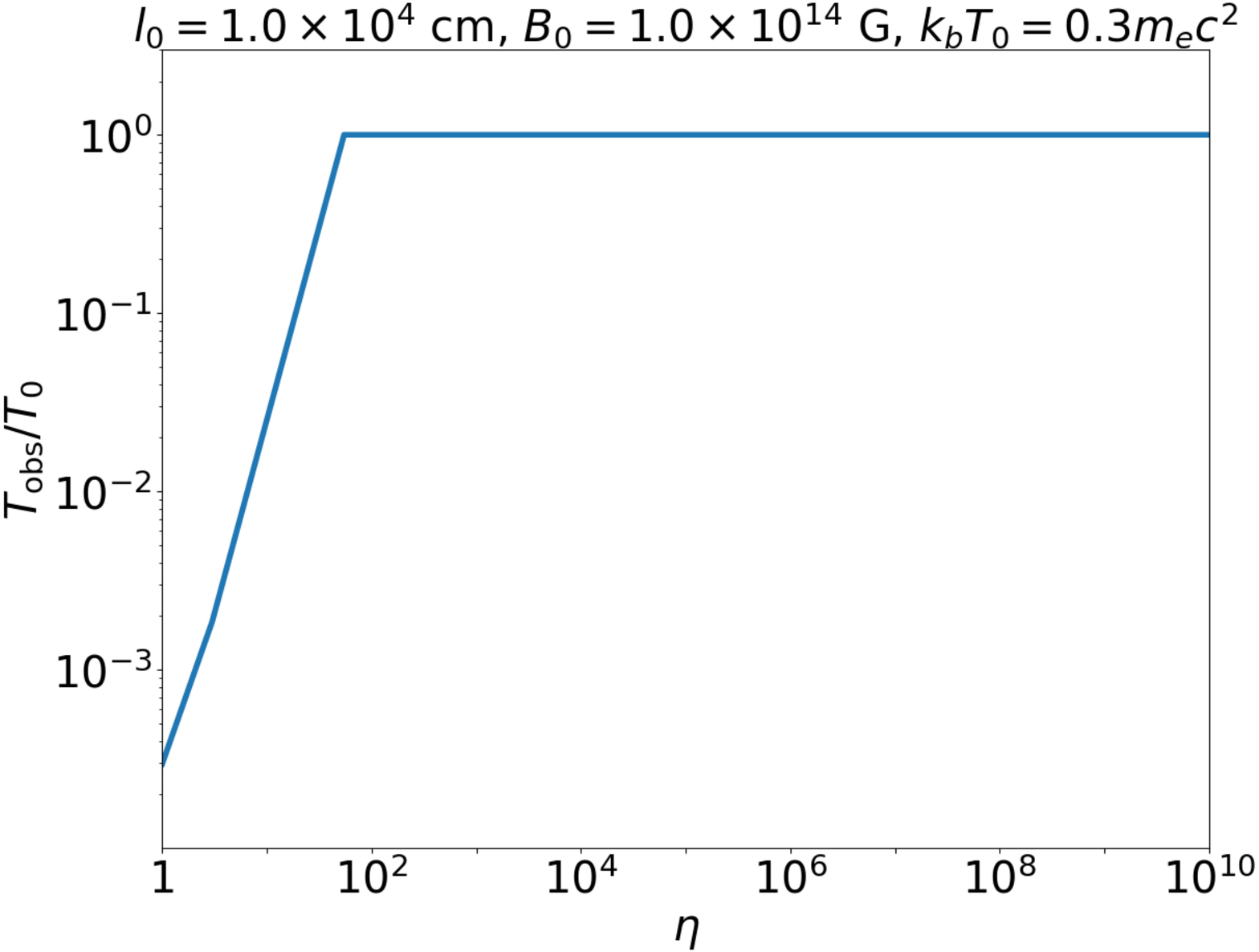}
\caption{
  $\eta$ dependence of the observed temperature (Equation~(\ref{eq:T_obs})) normalized by the initial temperature $T_0$.
  In the matter-dominated phase, $T_{\rm obs}$ decreases because the Lorentz factor cannot be higher than $\eta$ and the blue shift cannot compensate the decrease in the temperature due to expansion.
  For high $\eta$, we ignore the temperature drop due to the lateral expansion.
  The drop is so small that this simplification does not change the result significantly (see footnote~\ref{fn:l}).
\label{fig:typical_T}}
\end{figure}

Figure~\ref{fig:typical_T} shows $\eta$-dependence of the observed temperature $T_{\rm obs}=\Gamma T$ normalized by the initial temperature $T_0$.
In this figure, we ignore the decrease in temperature due to diffusion for $\eta>\eta_3$.
If we do not ignore this, the observed temperature for $\eta>\eta_3$ becomes a bit lower than the initial temperature unlike the spherical symmetric case \citep{Goo1986,Pac1986}.
This is because the fireball expand laterally, and thus the temperature in the comoving frame decreases.
The radial Lorentz factor does not change so much during the lateral expansion because $t_{\rm diff}$ is smaller than $t_{\rm dyn}$, and as a result, the observed temperature decreases.
However, we ignore this effect because this effect does not change the result by a factor of 2.
On the other hand, in the baryon-thick case, the observed temperature equals the initial temperature because diffusion does not occur in this case.
In baryon-dominated phase ($\eta<\eta_*$), the observed temperature drops from $T_0$.

\section{Application to FRB 20200428A}\label{sec:200428}\

\begin{figure*}
\includegraphics[width = \textwidth]{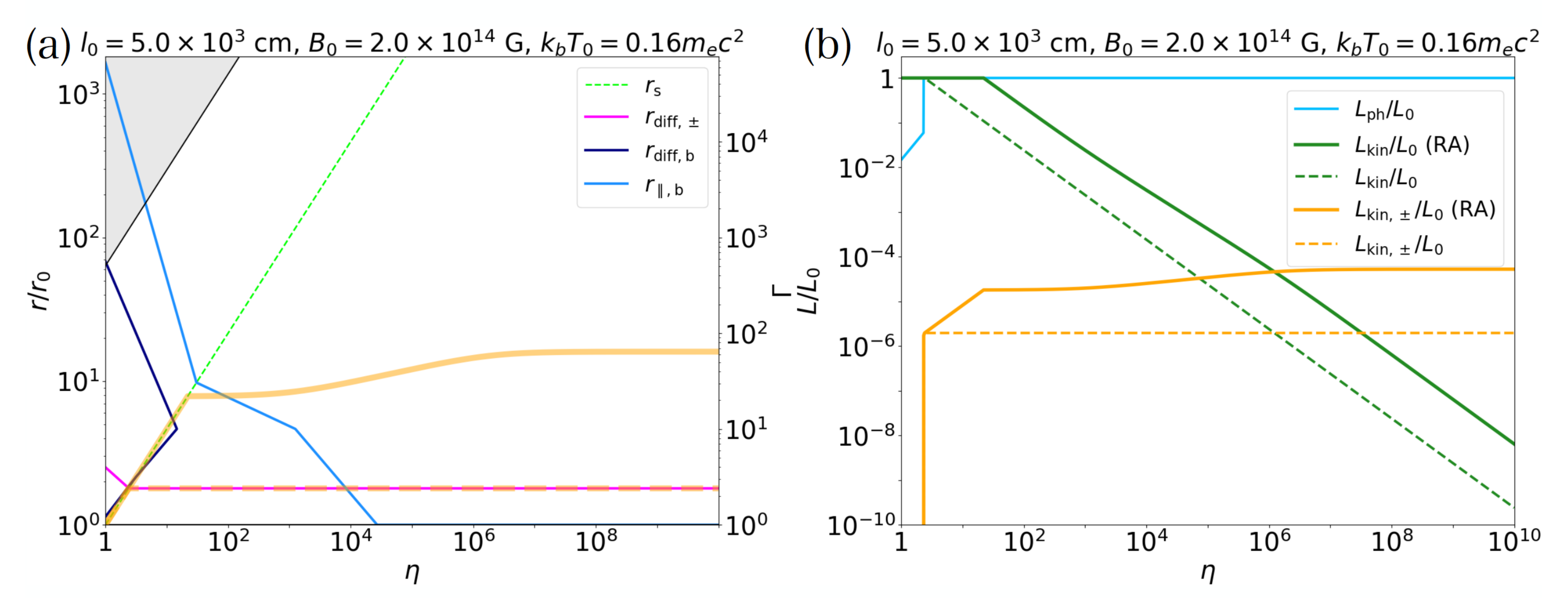}
\caption{
  (a)$\eta$ dependence of the saturation radius, $\tilde{r}_{\rm S}$ (green dashed line, Equation~(\ref{eq:r_s})), the lateral-diffusion radius for pair component, $\tilde{r}_{\rm diff,\pm}$ (pink, see Equation~(\ref{eq:diff_p})), the lateral-diffusion radius for baryonic electron component $\tilde{r}_{\rm diff, b}$ (navy, Equation~(\ref{eq:radius_db})), the photospheric radius for baryonic electron component $\tilde{r}_{\rm \parallel, b}$ (light blue, Equation~(\ref{eq:radius_pb}), scale on the left axis), the final Lorentz factor with radiative acceleration beyond the photospheric radius (yellow solid, scale on the right axis), and that without radiative acceleration beyond the photospheric radius (yellow dashed, scale on the right axis).
    For the yellow solid line, the left axis does not show the radius where the final Lorentz factor is realized.
The gray shaded region is above the Alfv\'{e}n radius where the assumption of expansion along the flux tube is violated.
  The lateral-diffusion radius for the baryonic electron component ($r_{\rm diff,b}$) increases as $\eta$ increase until the scattering-suppression radius becomes $\tilde{r}_{\rm E}\sim4$.
  (b) The photospheric luminosity (light blue), kinetic luminosity of the baryonic component (green), and that of the pair component (orange; see Equations~(\ref{eq:Lph_pp})--(\ref{eq:Lkine_pp}), and (\ref{eq:Lph_bd})--(\ref{eq:Lkine_bm})).
  Due to the suppression of the cross section for the E-mode photons, $\eta_3$ does not exist and there is no parameter region for the baryon-thick case (see Section~\ref{sec:blc}).
  }
\label{fig:200428}
\end{figure*}

\begin{figure}
\includegraphics[width=\linewidth]{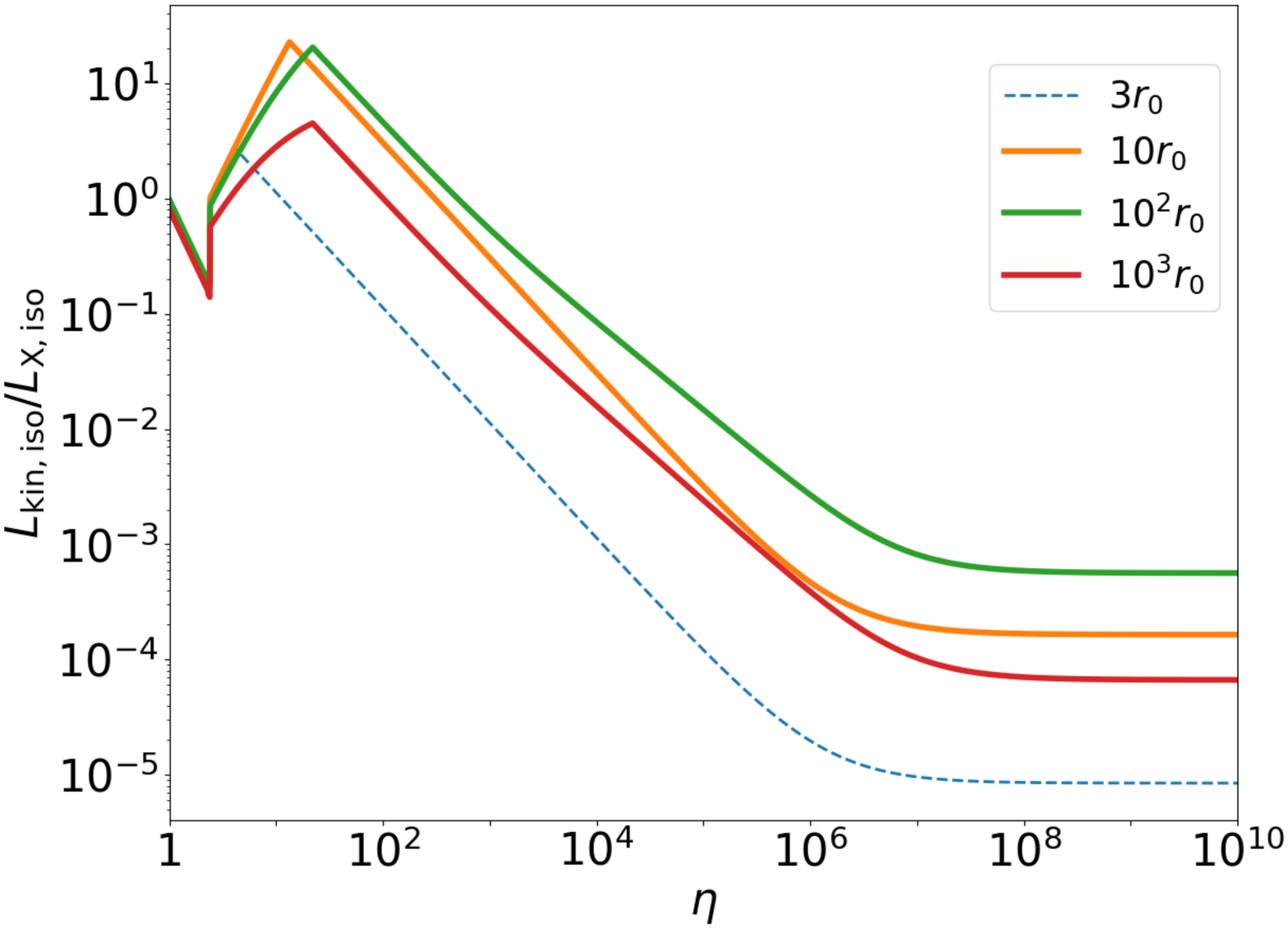}
\caption{
  Ratio of the maximum isotropic luminosity of the emitted FRB (Equation~(\ref{eq:isoradio})) to the isotropic luminosity of X-ray burst (Equation~(\ref{eq:isoX})).
  Each line shows the ratio for the FRB emitted at the radius in the graph legends, and the dashed line for $r=3r_0$ is for the reference to understand the $r$-dependence of the ratio. 
  For $\eta\lesssim 4$, the fireball is optically thick at $3r_0$ (blue dashed line) and $10r_0$ (orange line), and we omit $\eta\lesssim 4$ for these lines.
  Due to the radiative acceleration and relativistic beaming, the isotropic luminosity of FRB increases for small $r$.
  For large $r$, the expansion along the dipole magnetic field decreases the isotropic luminosity of FRB.
  \label{fig:Lratio}}
\end{figure}

We apply our fireball model to the short burst associated with FRB 20200428A.
All the five cases do not appear for the parameters adopted.
The adopted parameters are $\ell_0=5\times10^3\,{\rm cm}$, $B_{\rm Q}B_0=2\times 10^{14}\,{\rm G}$, and $m_{\rm e}c^2T_0=80\,{\rm keV}$.
These parameters are chosen based on the observations of X-ray burst associated with FRB 20200428A.
With these $T_0$ and $\ell_0$, in the pair-diffusion case, $T_{\rm obs}\sim80\,{\rm keV}$ and $L_{\rm ph,iso}\sim 10^{41}\,{\rm erg\,s^{-1}}$ are realized.
The observed X-ray short burst is realized by the photospheric emission of this expanding fireball for $\eta_*\lesssim\eta$.

Figure~\ref{fig:200428}(a) shows $\eta$-dependence of each radius.
In this case, $\eta_2$ and $\eta_3$ do not appear.
The initial flux tube is so thin that the diffusion timescale of the baryonic electron component is shorter than the dynamical timescale even for small $\eta$.
Thus, the fireball is diffusively thin in the lateral direction to the baryonic electron component for a wide range of $\eta$ values.
As a result, there is no intersection of lines for $\tilde{r}_{\rm diff,\pm}(\eta)=\tilde{r}_{\rm diff,b}(\eta)$ and $\tilde{r}_{\rm diff,b}(\eta)=\tilde{r}_{\rm \parallel,b}(\eta)$ in $\eta$-$\tilde{r}$ plane, and $\eta_2$ and $\eta_3$ do not appear.
For $\eta<\eta_{\rm b}\simeq 1\times10^6$, the kinetic energy of the bariyonic component is higher than that of the pair plasma (see Figure~\ref{fig:200428}(b) and Equation~(\ref{eq:eta_crit})).
For these values, the bayonic electron component enhances the kinetic luminosity of the outflow.

What are the implications of the fireball evolution for FRBs?
Among the outflow of the fireball, the kinetic energy of the pair and the baryonic component is available for powering the observed FRBs.
Thus, we compare the photospheric luminosity, $L_{\rm ph}$, which is the luminosity of the X-ray short burst, and the total kinetic luminosity of the pair and the baryonic component, $L_{\rm kin}+L_{\rm kin,\pm}$, which is the maximum luminosity of the FRB.
We do not focus on any specific emission mechanism of the FRB, which converts the kinetic energy of the outflow to the energy of the FRB.
It may be synchrotron maser \citep[e.g.,][]{MetMar2019}, emission by a bunch formed in the magnetosphere \citep[e.g.,][]{Uso1987}, or other mechanisms.

The observed isotropic luminosity of the FRB is three orders of magnitude lower than that of the X-ray short burst, and thus the isotropic kinetic luminosity must be higher than $10^{-3}$ times the luminosity of the photospheric emission.
Because the observed value is the isotropic luminosity (not the true luminosity in Figure~\ref{fig:200428}(b)), we compare the isotropic luminosity of the outflow.
Figure~\ref{fig:Lratio} shows the ratio of the maximum isotropic luminosity of the emitted FRB,
\begin{equation}
  L_{\rm kin,iso}(r)=\frac{4}{\left\{\theta_0\tilde{r}^{1/2}+\Gamma_{\rm RA}\left(r\right)^{-1}\right\}^2}\left(L_{\rm kin}(r)+L_{\rm kin,\pm}(r)\right),
  \label{eq:isoradio}
\end{equation}
where $r$ is the radius where FRB is emitted, to the isotropic luminosity of the X-ray burst,
\begin{equation}
  L_{\rm X,iso}=\frac{4}{\left\{\theta_0\tilde{r}_{\rm diff,\pm}^{1/2}+\Gamma\left(r_{\rm diff,\pm}\right)^{-1}\right\}^2}L_{\rm ph}.
    \label{eq:isoX}
\end{equation}
The beaming angle of the X-ray burst is determined at $\tilde{r}_{\rm diff,\pm}$ because we consider the radiation-dominated case and the photons escape from the fireball at $\tilde{r}_{\rm diff,\pm}$ for radiation-dominated case (see Figure~\ref{fig:200428}(a)).
The half-opening angle of the X-ray is determined at the radius where photons escape, and that of the FRB is determined by the radius where FRB is emitted.
Because of the radiative acceleration and the dipolar shape of the flow, the half-opening angle of the FRB, $\theta_0\tilde{r}^{1/2}+\Gamma^{-1}$, changes as the radius changes.
In order to understand $r$-dependence of $L_{\rm kin,iso}/L_{\rm X,iso}$, we consider $\eta\sim10^2$--$10^6$ as an example.
In this parameters, the baryonic component dominates kinetic luminosity.
For $r\lesssim30r_0$, the baryonic outflow is accelerated by the radiation and the half-opening angle, $\theta_0\tilde{r}^{1/2}+\Gamma^{-1}\sim\Gamma^{-1}$, decreases as the radius increases.
Thus, the isotropic luminosity of the emitted FRB increases as the radius increases due to the relativistic beaming.
For $r\gtrsim30r_0$, the radiative acceleration does not occur and the half-opening angle, $\theta_0\tilde{r}^{1/2}+\Gamma^{-1}\sim\theta_0\tilde{r}^{1/2}$, increases as the radius increase.
Thus, the isotropic luminosity of the emitted FRB decreases as the radius increases.
$\eta$-dependence of the ratio is a consequence of that of $L_{\rm kin}+L_{\rm kin,\pm}$.

If there is a sufficient baryonic component in the fireball, the kinetic energy of the outflow can be high enough to generate the observed FRB.
For example, if the FRB is emitted at $r=10^3r_0$, the dimensionless entropy should be lower than $\eta\sim10^6$ for $L_{\rm kin,iso}/L_{\rm X,iso}\gtrsim10^{-3}$ (Figure~\ref{fig:Lratio}).
For the pair fireball, the ratio of the luminosity can be as high as $\sim 6\times 10^{-4}$ if FRB is emitted at $r\sim10^2r_0$.
This value is close to the observed one within a factor of 2. 
Considering the approximations we made (e.g., constant energy injection, no entropy generation, no electromagnetic loading, black body spectrum in radiative acceleration), the pair fireball is still viable for explaining the observed FRBs if the radiation efficiency of the FRBs is high.

The ratio, $L_{\rm kin, iso}/L_{\rm X, iso}$, depends on the initial parameters of the expanding fireball and the magnetic field, $l_0,\,T_0$, and $B_0$.
For other X-ray short bursts, i.e., other values of $l_0,\,T_0$, and $B_0$,
the ratio, $L_{\rm kin, iso}/L_{\rm X, iso}$, can change.
This ratio changes by 1-2 orders of magnitude if, with fixing other parameters, one of $l_0,\,T_0$, and $B_0$ is changed in the range $l_0=1\times 10^3$--$10^5\,{\rm cm}$, $B_{\rm Q}B_0=10^{14}$--$2\times10^{15}\,{\rm G}$, and $T_0= 0.16$--$0.3$.

Without the pair annihilation of the positrons with the electrons associated with baryons, the number density of the electrons and positrons in the lab frame in the optically thin regime is
\begin{align}
n_{\rm L}(r)&=\Gamma\left(2n_+ +\frac{\rho}{m_{\rm p}}\right)\nn\\
&=\left[\frac{1}{\sigma_{\rm T}r_0}\left(\frac{\sigma(T_\parallel,B_\parallel)}{\sigma_{\rm T}}\right)^{-1}\tilde{r}_{\rm \parallel}^5+\left(\frac{\pi^2m_{\rm e}^4c^3}{15\hbar^3m_{\rm p}}\right)T_0^4\eta^{-1}\right]\tilde{r}^{-3},
\label{eq:number_prop}
\end{align}
where $T_\parallel$ and $B_\parallel$ are the temperature and the magnetic field at $r=r_{\rm \parallel}$, and we have used Equation~(\ref{eq:baryon_dyn}), optically thinning condition for pairs, $2n_{+,\parallel}\sigma(T_\parallel,B_\parallel) r_0\tilde{r}_\parallel/\Gamma_{\parallel}=1$, and the number density flux conservation without pair annihilation, $n_+\Gamma r^3={\rm const.}$, in $r>r_\parallel$.
The first term of Equation~(\ref{eq:number_prop}) is the number density of pair plasma and the second is that of baryonic electron component.
For the pair-diffusion case, the first term is dominant.
For the pair-diffusion-baryon-photosphere case, only the first term contributes in the newly created pair plasma fireball (the yellow pair fireball in Figure~\ref{fig:overview}), where baryons do not exist, and the only second contributes in the initial flux tube where pair annihilation occurs.
For the baryon-diffusion case, the baryon-photosphere case, and the baryon-dominated case, the second term is dominant because the baryon number density is higher than that of the pairs at the radius where photons begin to escape.
In deriving the second term, optically thinning condition for pairs is not used and thus this term is valid in these three cases.
Compared to the Goldreich-Julian density \citep{GolJul1969}, $n_{\rm GJ}(r)$, this number density is 
\begin{align}
\frac{n_{\rm L}(r)}{n_{\rm GJ}(r)}
&\sim 10^{7}\left[\left(\frac{\sigma(T_\parallel,B_\parallel)}{\sigma_{\rm T}}\right)^{-1}\left(\frac{\tilde{r}_{\rm \parallel}}{2}\right)^5 +\left(\frac{T_0}{0.16}\right)^4\left(\frac{\eta}{8.5\times10^4}\right)^{-1}\right]\nn\\
&\times\left(\frac{B_{\rm Q}B}{2\times10^{14}\,{\rm G}}\right)\left(\frac{P}{3\,{\rm s}}\right)^{-1},
\end{align}
times higher, where $P$ is the period of the magnetar.
For the pair-diffusion case with the parameters in Figure~\ref{fig:200428}, $\sigma(T_\parallel,B_\parallel)$ is about $6\times10^{-2}\sigma_{\rm T}$.
Whether the fireball particles scatter and prevent FRB photons from escaping the magnetosphere or not is a controversial issue \citep[e.g.,][]{Iok2020,Lyu2021,Bel2021,Bel2022, QuKum2022}.

\section{Conclusion and discussion}\label{sec:summary}
In this paper, we investigate the fireball evolution in a magnetic flux tube.
We construct a steady-state solution of the fireball expanding along dipole magnetic field lines (Section~\ref{fluid}).
We overview the effect of the magnetic field on the fireball.
The cross section for an E-mode photon is suppressed, which makes it easy for the photons in the fireball to escape (Section~\ref{sec:magnetic}).
Using these results, we investigate the escape process of the photon from the fireball and reveal the importance of the lateral-diffusion from the flux tube (Section~\ref{sec:beaking}).
This is because the fireball confined in a flux tube may be narrow, and the diffusion of photons in the direction lateral to the flux tube can be the main process of the photon escape.
This is a major difference from the fireball of gamma-ray bursts.
We reveal the dependence of the lateral-diffusion radius and the photospheric radius on the dimensionless entropy, $\eta$, taking into account the effect of the strong magnetic field.
We also consider the effects of baryon for the fireball evolution in a flux tube for the first time.
The baryonic electron component can keep the fireball optically and diffusively thick until it is accelerated to a higher Lorentz factor than the pure pair plasma fireball.
Due to the lateral-diffusion, the fireball behaves more diversely than the fireball of gamma-ray bursts.
The radiative acceleration beyond the photospheric radius via the resonant scattering is also investigated in the tube fireball for the first time.
We classify the behaviors into five cases and evaluate the photospheric luminosity, kinetic luminosity of the baryonic component, and that of the pair plasma (Section~\ref{sec:cases}).
The behavior of these luminosities shows different dependence on $\eta$ than the fireball of gamma-ray bursts.

We apply our model to the X-ray short burst from a magnetar associated with the Galactic FRB, FRB 20200428A (Section~\ref{sec:200428}).
If the dimensionless entropy is lower than $\sim 10^5$--$10^6$,
the isotropic kinetic energy of the fireball can be high enough to explain the isotropic energy of the observed FRB.
A pair fireball is still viable for explaining the FRB energy (with a factor of 2 uncertainty) thanks to the radiative acceleration beyond the photospheric radius via resonant scattering.

In this paper, we neglect some effects that can change the dynamics or the observed spectrum.
If the energy of the magnetic field is loaded into the fireball, it can be an additional energy source to accelerate the fireball \citep{Dre2002,DreSpr2002,TanTom2020,Lyu2022,BarSha2022}.
The decay of the Alfv{\'e}n wave at high latitude also can inject energy into the fireball.
If the diffusion occurs, there may be a strong pressure gradient at the lateral-diffusion radius.
This pressure gradient can accelerate the fireball \citep{TanTom2020}.

If the resonant scattering occurs, the observed radiation spectrum may become different from the black body spectrum \citep[e.g.][]{Lyu1986,Lyu2002,Bel2013,YamLyu2020}.
This is because the photons with different energies escape from different radii, that is, the observed spectrum is multi-color.
The resonant scattering also can change the spectral shape from the initial black body.
The deviation from the black body spectrum would change the final Lorentz factor (Equation~(\ref{eq:Gamma_res})), which is derived assuming the black body spectrum.
We need a numerical calculation to reveal the spectrum shape and its back reaction on the radiative acceleration.
This is future work to consider.

The outflow with high kinetic luminosity can create nebula emission \citep[e.g.,][]{Lyu2014,MurKas2016,Bel2017}.
If baryon is loaded, the kinetic luminosity of the outflow can be high and the nebula could be bright.
It may be bright as a possible persistent radio counterpart.

We also do not consider the detailed emission mechanism of the FRB.
While the emission mechanism of the FRB is not known, synchrotron maser instability at a shock and coherent bunch emission in a magnetosphere are possible candidates \citep[][]{Lyu2021_emission}.
If the energy injection from the base of the fireball varies with time the final Lorentz factor also varies with time, and might cause an internal shock in the outflow.
Also, shocks may be generated outside the magnetosphere by the outflow.
These shocks might be responsible for the FRB emission.

If FRB occurs in a binary system, as the observed periodicity might suggest \citep{LyuBar2020,IokZha2020,WadIok2021,BarPop2022}, there may be a stellar wind from the companion.
The collision of the fireball outflow and the stellar wind might create a possible counterpart of the FRB.
The collision also might broaden the funnel size of the binary comb model, and affect the periodicity.

\section*{Data Availability}
The data underlying this article will be shared on reasonable request to the corresponding author.

\section*{Acknowledgements}
We are grateful to Kazuya Takahashi, Hamidani Hamid, Wataru Ishizaki, Koutarou Kyutoku,
Katsuaki Asano, Kyohei Kawaguchi, Tomohisa Kawashima, Shigeo S. Kimura, Tomoya Kinugawa, Pawan Kumar, Kohta Murase, Kenji Toma  for fruitful discussion and valuable comments.
We also thank the participants and the organizers of the YITP workshop YITP-W-22-18, "Fast Radio Bursts and Cosmic Transients", for their generous support and helpful comments.
This work is supported by Grants-in-Aid for Scientific Research No. 20J13806, 22K20366 (TW), 22H00130, 20H01901, 20H01904, 20H00158, 18H01215, 17H06357, and 17H06362 (KI) from the Ministry of Education, Culture, Sports, Science and Technology (MEXT) of Japan.



\bibliographystyle{mnras}
\bibliography{cite}{}


\appendix
\section{Multipolar Magnetic Fields}\label{sec:multipole}
In this appendix, we consider the dynamics along multipolar field lines.
For $2^{n-2}$ pole magnetic field, the strength of the magnetic field is 
\begin{equation}
    B(r)=B_0\tilde{r}^{-n}.
    \label{eq:Br_m}
\end{equation}
Here, we neglect $\vartheta$ dependence .
The approximation using the the radial coordinate, $r$, to measure the position of the fluid breaks down at $\tilde{r}=\theta_0^{-2/(n-2)}$.

For $2^{n-2}$ pole magnetic field, $r^2\Delta \Omega\propto r^{n}$ along the magnetic field lines.
Using Equations~(\ref{eq:baryon_dyn})--(\ref{eq:entropy_dyn}), $r$-dependences of $e$, $\Gamma$, and $\rho$ are
\begin{eqnarray}
\tilde{e}(r)&=&
\begin{dcases}
\tilde{r}^{-2n} &({\rm RD})\\
\eta^{-4/3}\tilde{r}^{-4n/3}&({\rm MD}),
\end{dcases}\label{eq:sol_e_m}\\
\Gamma(r)&=&
\begin{dcases}
\tilde{r}^{n/2}&({\rm RD})\\
\eta&({\rm MD}),
\end{dcases}\label{eq:sol_gamma_m}\\
\tilde{\rho}(r)&=&
\begin{dcases}
\tilde{r}^{-3n/2}&({\rm RD})\\
\eta^{-1}\tilde{r}^{-n}&({\rm MD}).
\end{dcases}\label{eq:sol_rho_m}
\end{eqnarray}
The fluid becomes baryon dominant at $\tilde{r}_{\rm S}\coloneqq\eta^{2/n}$.
The Alfv\'{e}n radius where the energy density of the fluid becomes higher than the magnetic energy density is
\begin{equation}
\tilde{r}_{\rm A}=\left(\frac{15}{8\pi^3\alpha}\right)^{1/n}B_0^{2/n}T_0^{-4/n}\eta^{2/n}.
\end{equation}
The dependences, $\tilde{r}_{\rm B}$, $\tilde{r}_{\rm S}$, and $\tilde{r}_{\rm A}$ are summarized in Table~\ref{tab:radbar_summary_m}.

In the multipolar case, the diffusion radius and the photospheric radius is determined by the same procedure discussed in this paper.
$\gamma$--$\zeta$ in Equation~(\ref{eq:tau_b}) and (\ref{eq:diff_b}) are modified as in Table~\ref{tab:index_baryon_m}.
$\gamma^\prime$--$\zeta^\prime$ and $A$ in Equation~(\ref{eq:tau_p}) and (\ref{eq:diff_p}) are also modified as in Table~\ref{tab:index_pair_m}.

\begin{table*}
 \caption{Summary of the evolution of physical quantities and the characteristic radii for multipolar magnetic field.}
  \label{tab:radbar_summary_m}
 \centering
  \begin{tabular}{ccc}
   \hline
     & Radiation-dominated & Baryon-dominated \\
   \hline \hline
   $\tilde{T}$ & $\tilde{r}^{-n/2}$ &$\eta^{-1/3}\tilde{r}^{-n/3}$\\
   $\Gamma$ & $\tilde{r}^{n/2}$ & $\eta$\\
   $\tilde{\rho}$ & $\tilde{r}^{-3n/2}$ & $\eta^{-1}\tilde{r}^{-n}$\\
    \hline    
   $\tilde{r}_{\rm S}$ & $\eta^{2/n}$ & $\eta^{2/n}$ \\
   \hline
    $\tilde{r}_{\rm A}$ &no solution & $\left(15/8\pi^3\alpha\right)^{1/n}B_0^{2/n}T_0^{-4/n}\eta^{2/n}$\\
    \hline
   $\tilde{r}_{\rm B}$ ($B\gg B_{\rm Q}$) & no solution  & $ \left(2B_0\right)^{3/n}T_0^{-6/n}\eta^{2/n}$ \\
   $\tilde{r}_{\rm B}$ ($B\ll B_{\rm Q}$) & $B_{0}^{2 / n}T_{0}^{-2 / n}$ & $B_{0}^{3 / 2n}T_{0}^{-3 / 2n} \eta^{1/2n}$\\
   \hline
   $\tilde{r}_{\rm E}$ & $\left(4\pi^2/5\right)^{-1/n}T_0^{-2/n}B_0^{2/n}$ & $\left(4\pi^2/5\right)^{-3/4n}T_0^{-3/2n}B_0^{3/2n}\eta^{1/2n}$\\
   \hline
  \end{tabular}
\end{table*}

\begin{table*}
 \caption{Variables in Equations~(\ref{eq:tau_b}) and (\ref{eq:diff_b}) for multipolar magnetic field.
 The definition of each case are the same as Table~\ref{tab:index_baryon}.}
 \label{tab:index_baryon_m}
 \centering
  \begin{tabular}{c|ccccccc}
   \hline
    & $\tau_{\rm b0}$&$\gamma$ & $\delta$ & $\epsilon$ & $\zeta$\\
   \hline \hline
   RD/O-mode ($\tilde{r}_{\rm S}>\tilde{r}>\tilde{r}_{\rm E}$) & $e_0\sigma_{\rm T} r_0/m_{\rm p}c^2$& $-1$ & $-2n+1$ & $-1$ & $-1$\\
   \hline
     MD/O-mode ($\tilde{r}>\tilde{r}_{\rm S}$, $\tilde{r}_{\rm E}$)& $e_0\sigma_{\rm T} r_0/m_{\rm p}c^2$& $-3$ & $-n+1$ & $-1$ & $-1$\\
   \hline
   RD/E-mode ($\tilde{r}_{\rm E},\tilde{r}_{\rm S}>\tilde{r}$) & $(4\pi^2T_0^2B_0^{-2}/5)e_0\sigma_{\rm T} r_0/m_{\rm p}c^2$& $-1$ & $-n+1$ & $-1$ & $n-1$\\
   \hline
   MD/E-mode ($\tilde{r}_{\rm S}>\tilde{r}>\tilde{r}_{\rm E}$) &$(4\pi^2T_0^2B_0^{-2}/5)e_0\sigma_{\rm T} r_0/m_{\rm p}c^2$ & $-11/3$ & $n/3+1$ &  $-5/3$ & $4n/3-1$\\
   \hline
  \end{tabular}
\end{table*}

\begin{table*}[htb]
 \caption{Variables in Equations~(\ref{eq:tau_p}) and (\ref{eq:diff_p}) for multipolar magnetic field.
 The definition of each case are the same as Table~\ref{tab:index_pair}.}
 \label{tab:index_pair_m}
 \centering
  \begin{tabular}{c|cccccccc}
   \hline
    & $\tau_{\rm \pm0}$&$\gamma^\prime$ & $\delta^\prime$ & $\epsilon^\prime$ & $\zeta^\prime$ & $A$\\
   \hline \hline
   RD/O-mode/lL & $n_\pm(T_0,B_0)\sigma_{\rm T}r_0$& $0$ & $-7n/4+1$  & $0$ & $n/4-1$ & $\tilde{r}^{n/2}-1$\\
   \hline
   RD/O-mode/hL & $n_\pm(T_0)\sigma_{\rm T}r_0$& $0$ & $-5n/4+1$ & $0$ & $3n/4-1$ & $\tilde{r}^{n/2}-1$\\
   \hline
   MD/O-mode/lL & $n_\pm(T_0,B_0)\sigma_{\rm T}r_0$ & $-7/6$ & $-7n/6+1$& $5/6$ & $-n/6-1$&$\eta^{1/3}\tilde{r}^{n/3}-1$\\
   \hline
   MD/O-mode/hL & $n_\pm(T_0)\sigma_{\rm T}r_0$& $-3/2$ & $-n/2+1$ & $1/2$ & $n/2-1$ &$\eta^{1/3}\tilde{r}^{n/3}-1$\\
   \hline
   RD/E-mode/lL & $(4\pi^2T_0^2B_0^{-2}/5)n_\pm(T_0,B_0)\sigma_{\rm T}r_0$& $0$ & $-3n/4+1$ & $0$ & $5n/4-1$ & $\tilde{r}^{n/2}-1$\\
   \hline
   MD/E-mode/lL & $(4\pi^2T_0^2B_0^{-2}/5)n_\pm(T_0,B_0)\sigma_{\rm T}r_0$& $-11/6$ & $n/6+1$ & $1/6$ & $7n/6-1$&$\eta^{1/3}\tilde{r}^{n/3}-1$\\
   \hline
  \end{tabular}
\end{table*}

\section{Pair-photosphere case}\label{sec:thin}
In this section, we consider the case of $\tilde{r}_{\rm diff,\pm}\geq\tilde{r}_{\rm \parallel,\pm}$.
This case occurs for a high initial temperature case ($\tilde{T}\gtrsim m_{\rm e}c^2$ for $\ell_0=10^4\,{\rm cm}$) or a large initial fireball size case ($\ell_0\gtrsim10^5\,{\rm cm}$ for $\tilde{T}\sim0.3m_{\rm e}c^2$).

In this case, the diffusion does not occur for both baryonic electron components and pair plasma.
Thus, there are only three cases, pair-photosphere case, baryon-photosphere case, and baryon-dominant case, as in the spherically symmetric fireball.
In the pair-photosphere case, the baryonic electron component does not contribute to the optical depth, and at the photosphere for pairs, the photons and the outflow are radiated.
Thus, all the luminosities are determined at $\tilde{r}=\tilde{r}_{\rm \parallel,\pm}$
\begin{align}
L_{\rm ph}&=\pi \ell^2(r_{\rm \parallel,\pm})caT^4(r_{\rm\parallel,\pm})\Gamma(r_{\rm \parallel,\pm})^2\\
&=L_0\nn\\
L_{\rm kin}&=\pi \ell^2(r_{\rm \parallel,\pm})\rho(r_{\rm \parallel,\pm})c^3\Gamma(r_{\rm \parallel,\pm})\Gamma_{\rm RA}(r_{\rm \parallel,\pm}),\\
&\simeq \begin{cases}
    L_{0} f_{\rm RA}\eta^{-1}\tilde{r}_{\parallel, \pm}^{1 / 2} &(\Gamma_{\rm RA}(r_{\rm \parallel,\pm})<\eta)\nn\\
    L_0 &(\Gamma_{\rm RA}(r_{\rm \parallel,\pm})\geq\eta)
    \end{cases},\nn\\
L_{\rm kin,\pm}&=\pi \ell^2(r_{\rm \parallel,\pm}) m_{\rm e}c^3n_{\rm e}\Gamma(r_{\rm \parallel,\pm})\Gamma_{\rm RA}(r_{\rm \parallel,\pm})\\
  &\sim 2\times10^{-5}L_0f_{\rm RA,1.5}\theta_{\rm 0, -2}^2L_{0,40}^{-1}r_{\rm \parallel,\pm,0.5}^{11/2}\nn\\
  &\times\left(\frac{\sigma_{\rm T}}{\sigma(T(r_{\parallel,\pm}),B(r_{\rm \parallel,\pm}))}\right)\nn.
\end{align}

In the baryon-photosphere case, the baryonic electron component determines the photospheric radius, and thus the luminosities are determined at this radius.
The luminosities are determined at $\tilde{r}=\tilde{r}_{\rm \parallel,b}$, and the expression is the same as Equations~(\ref{eq:Lph_bt})--(\ref{eq:Lkine_bt}).
In the baryon-dominant case, the luminosity is expressed in the same formula as in Equation~(\ref{eq:Lph_bt})-(\ref{eq:Lkine_bt}).


\bsp	
\label{lastpage}
\end{document}